\documentclass[twocolumn]{aastex61}
\usepackage{color}
\usepackage{bm}
\usepackage[normalem]{ulem}

\usepackage{hyperref}

\usepackage{natbib}
\begin{document}

\title{The APOGEE-2 Survey of the Orion Star Forming Complex: \\ I. Target Selection and Validation with early observations} 

\author{J'Neil Cottle}
\affil{Department of Physics \& Astronomy, Western Washington University, Bellingham WA 98225-9164 USA} 

\author{Kevin R. Covey}
\affil{Department of Physics \& Astronomy, Western Washington University, Bellingham WA 98225-9164 USA} 
\email{kevin.covey@wwu.edu}

\author{Genaro Su\'arez}
\affil{Instituto de Astronom\'ia, Universidad Nacional Aut\'onoma de M\'exico, Unidad Acad\'emica en Ensenada, Ensenada BC, 22860 Mexico}

\author{Carlos Rom\'an-Z\'u\~niga}
\affil{Instituto de Astronom\'ia, Universidad Nacional Aut\'onoma de M\'exico, Unidad Acad\'emica en Ensenada, Ensenada BC, 22860 Mexico}

\author{Edward Schlafly}
\affil{Max-Planck-Institut f\"ur Astronomie, K\"onigstuhl 17, D-69117 Heidelberg, Germany}

\author{Juan Jose Downes}
\affil{Instituto de Astronom\'ia, Universidad Nacional Aut\'onoma de M\'exico, Unidad Acad\'emica en Ensenada, Ensenada BC, 22860 Mexico}
\affil{Centro de Investigaciones de Astronom\'{\i}a, AP 264, M\'erida 5101-A, Venezuela}

\author{Jason E. Ybarra}
\affil{Instituto de Astronom\'ia, Universidad Nacional Aut\'onoma de M\'exico, Unidad Acad\'emica en Ensenada, Ensenada BC, 22860 Mexico}
\affil{Department of Physics, Bridgewater College, 402 E College St, Bridgewater, VA 22812}

\author{Jesus Hernandez}
\affil{Instituto de Astronom\'ia, Universidad Nacional Aut\'onoma de M\'exico, Unidad Acad\'emica en Ensenada, Ensenada BC, 22860 Mexico}

\author{Keivan Stassun}
\affil{Department of Physics \& Astronomy, Vanderbilt University, Nashville, TN 37235, USA}

\author{Guy S. Stringfellow}
\affil{Center for Astrophysics and Space Astronomy, University of Colorado, UCB 389, Boulder, CO 80309, USA}

\author{Konstantin Getman}
\affil{Department of Astronomy \& Astrophysics, 525 Davey Laboratory, Pennsylvania State University, University Park, PA 16802, USA}

\author{Eric Feigelson}
\affil{Department of Astronomy \& Astrophysics, 525 Davey Laboratory, Pennsylvania State University, University Park, PA 16802, USA}

\author{Jura Borissova}
\affil{Instituto de F\'isica y Astronom\'ia, Facultad de Ciencias, Universidad de Valpara\'iso, Av. Gran Breta\~na 1111, Playa Ancha, Casilla 5030, Valapara\'iso, Chile}
\affil{Millenium Institute of Astrophysics, Av. Vicu\~na Mackenna 4860, 7820436, Macul, Santiago, Chile}

\author{J. Serena Kim}
\affil{Steward Observatory, University of Arizona, Tucson, AZ 85721, USA}

\author{A. Roman-Lopes}
\affil{Department of Physics, Universidad de La Serena, Cisternas, 1200 La Serena, Chile}

\author{Nicola Da Rio}
\affil{Department of Astronomy, University of Virginia, Charlottesville, VA 22904-4325, USA}

\author{Nathan De Lee}
\affil{Department of Physics, Geology and Engineering Tech, Northern Kentucky University, Highland Heights, KY 41099, USA}
\affil{Department of Physics \& Astronomy, Vanderbilt University, Nashville, TN 37235, USA}

\author{Peter M. Frinchaboy}
\affil{Department of Physics and Astronomy, Texas Christian University, Forth Worth, TX 76129, USA}

\author{Marina Kounkel}
\affil{Department of Physics \& Astronomy, Western Washington University, Bellingham WA 98225-9164 USA} 

\author{Steven R. Majewski}
\affil{Department of Astronomy, University of Virginia, Charlottesville, VA 22904-4325, USA}

\author{Ronald E. Mennickent}
\affil{Universidad de Concepci\'on, Departmento de Astronom\'ia, Casilla 160-C, Concepci\'on, Chile}

\author{David L. Nidever}
\affil{Department of Physics, Montana State University, Bozeman, MT 59717, USA}

\author{Christian Nitschelm}
\affil{Unidad de Astronom\'ia, Universidad de Antofagasta, Av. Angamos 601, Antofagasta 1270300 Chile}

\author{Kaike Pan}
\affil{Apache Point Observatory and New Mexico University, P.O. Box 59, Sunspot, NM 88349, USA}

\author{Matthew Shetrone}
\affil{University of Texas at Austin, McDonald Observatory, Fort Davis, TX 79734, USA}

\author{Gail Zasowski}
\affil{University of Utah, Dept. of Physics \& Astronomy, 115 South 1400 East, JFB 201, Salt Lake City, UT 84112, USA}

\author{Ken Chambers}
\affil{Institute for Astronomy, University of Hawai'i, 2680 Woodlawn Drive, Honolulu, HI 96822, USA}

\author{Eugene Magnier}
\affil{Institute for Astronomy, University of Hawai'i, 2680 Woodlawn Drive, Honolulu, HI 96822, USA}

\author{Jeff Valenti}
\affil{Space Telescope Science Institute, Baltimore, MD, 21218, USA}

%\clearpage
\begin{abstract}
The Orion Star Forming Complex (OSFC) is a central target for the APOGEE-2 Young Cluster Survey.  Existing membership catalogs span limited portions of the OSFC, reflecting the difficulty of selecting targets homogeneously across this extended, highly structured region. We have used data from wide field photometric surveys to produce a less biased parent sample of young stellar objects (YSOs) with infrared (IR) excesses indicative of warm circumstellar material or photometric variability at optical wavelengths across the full 420 square degrees extent of the OSFC.  When restricted to YSO candidates with $H < 12.4$, to ensure S/N $\sim 100$ for a  six visit source, this uniformly selected sample includes 1307 IR excess sources selected using criteria vetted by Koenig \& Liesawitz and 990 optical variables identified in the Pan-STARRS1 3$\pi$ survey: 319 sources exhibit both optical variability and evidence of circumstellar disks through IR excess.  Objects from this uniformly selected sample received the highest priority for targeting, but required fewer than half of the fibers on each APOGEE-2 plate. We fill the remaining fibers with previously confirmed and new color-magnitude selected candidate OSFC members. Radial velocity measurements from APOGEE-1 and new APOGEE-2 observations taken in the survey's first year indicate that $\sim$90\% of the uniformly selected targets have radial velocities consistent with Orion membership.The APOGEE-2 Orion survey will include $>$1100 bona fide YSOs whose uniform selection function will provide a robust sample for comparative analyses of the stellar populations and properties across all sub-regions of Orion.
\end{abstract}

\section{Introduction}

Star forming regions are invaluable astrophysical laboratories. The pre-main sequence binaries within these regions enable stringent tests of stellar evolutionary models \citep[e.g., ][]{Stassun2014}, and the ages and kinematics of the full cluster population constrain the physical processes during the formation and early evolution of stars and clusters \citep[e.g., ][]{DaRio2014}.  Surveys of cluster populations with high-resolution, multi-object spectrographs provide the most efficient route to the precise measurements of stellar and kinematic properties that these tests require. Surveys conducted at optical wavelengths have provided important constraints on the membership, star formation histories, and dynamical states of the optically accessible members of relatively compact clusters \citep{Tobin2009,Jeffries2014,Sacco2015}, but investigations of embedded  or extended complexes have thus far been limited to smaller subsets of low-extinction \citep{Rigliaco2016} or centrally concentrated sources. 

The APOGEE (Apache Point Observatory Galactic Evolution Experiment) survey provides a unique opportunity to study the most embedded and extended regions of active star formation. The SDSS-IV APOGEE-2 survey aims to obtain infrared spectra of hundreds of thousands of red giant stars, in all components of the Milky Way, in both hemispheres, to reconstruct the Galaxy's star formation history \citep{Zasowski2013, Majewski2015}. The survey includes several additional `Goal Programs' that are not included in the APOGEE-2 survey's formal science requirements but address other areas of scientific interest, including star formation. The APOGEE spectrographs' infrared (1.51-1.7 $\mu$m) sensitivity, multiplex capability, and wide fields-of-view enable efficient surveys of extended and embedded (A$_V \geq$ 2-3 mag) sites of active star formation. 

The APOGEE-1 INfrared Survey of Young Nebulous Clusters (IN-SYNC), an SDSS-III ancillary program, provided a first demonstration of the precise stellar parameters \citep[$\sigma_{Teff} \sim$ 80 K; $\sigma_{log g} \sim$ 0.1 dex;][]{Cottaar2014} that APOGEE can provide for even the youngest stars.  The IN-SYNC program obtained APOGEE spectra for $\sim$3500 Young Stellar Objects (YSOs) in the Orion A, Perseus, and NGC 2264 star forming regions, demonstrating that, even in the presence of substantial rotational broadening and spot-induced line profile variations, APOGEE spectra enable the identification of pre-main sequence binaries \citep{Fernandez2017} and provide reliable measurements of the radial velocities (RVs) of individual YSOs ($\sigma \sim$ 0.2-0.5 km/s) and the velocity dispersions of the bulk cluster populations \citep[$\sigma \sim$ 0.1-0.2 km/s; ][]{Foster2015,Cottaar2015}. APOGEE-1 observations within the Orion A cloud \citep{DaRio2016,DaRio2017} also demonstrate that joint analyses of stellar properties and kinematics can help untangle the physical structure and star formation history of a physically extended region with a broader range of stellar ages. \citet{DaRio2016} demonstrate, for example, that isochronal ages inferred from APOGEE-based T$_{eff}$ values are well correlated with several independent proxies for YSO evolutionary state, such as log $g$, total line-of-sight extinction, and mid-infrared SED slope.  Stellar age estimates derived from APOGEE spectra could be quite useful for measuring the timescales of various star formation processes, such as changes in the structure and composition of circumstellar disks \citep[e.g.,][]{Haisch2001, Andrews2005, Hernandez2007}.

The APOGEE-2 Young Cluster Survey aims to provide a homogeneous, high-quality catalog of RVs and stellar parameters for several nearby ($d <$ 1 kpc), young ($t <$ 125 Myr) clusters and star forming regions.  Measurements of the global velocity dispersions of these clusters, particularly at the largest cluster radii which are uniquely available from the APOGEE spectrograph's field of view, will enable new tests of the mechanisms by which young clusters form, thermalize, and often disperse \citep[e.g.,][]{Stutz2016}. Consistent measurements of precise stellar properties, particularly (model-dependent) stellar ages, within and across cluster environments, will help distinguish between models that predict fast \citep[e.g.][]{Elmegreen2000} or slow \citep[e.g.][]{Krumholz2007} timescales for the star formation process. 

The Orion Star Forming Complex (OSFC) is a particularly important target for the survey: this region is a critical benchmark for studies of low- and high-mass star formation, but the region's large angular extent ($>$500 deg$^2$) has typically limited observers to study isolated sub-populations. APOGEE-2's unique ability to efficiently observe thousands of YSOs across hundreds of square degrees, and place heavily obscured YSOs on an equal footing with optically visible sources, provides the opportunity to assemble the first self-consistent diagnosis of the global kinematics and star formation histories of this rich star forming region. This effort builds on the IN-SYNC survey of the Orion A molecular cloud, led by \citet{DaRio2016, DaRio2017}, by expanding the footprint of the survey to enable comparative studies of all the major sub-populations of this extended star forming region: Orion B, $\lambda$ Ori, $\sigma$ Ori, and Orion OB1. Selecting targets across this extended region is no simple task, however: existing membership catalogs typically span only a portion of the region targeted by this program, and merging members from multiple, disjoint literature catalogs would produce a complex selection function that would be difficult to account for in any analysis that seeks to compare the populations of YSOs located in different sub-regions of the OSFC. Only by selecting targets in a consistent and straightforward manner across the full survey footprint can we be confident that any differences detected in stellar kinematics, ages, or other stellar properties reflect intrinsic differences in the underlying population, rather than differences arising from selection biases in the underlying catalogs.

For example, consider the biases that may result from relying exclusively on a common youth indicator, such as the presence of an IR excess due to circumstellar material.  YSOs with IR excesses can be identified and classified into broad evolutionary categories \citep[i.e., Class 0/I/II/III; see ][]{Adams1987, Whitney2003} with mid-IR photometry, which is now widely available from wide-field surveys conducted by the \textit{Spitzer} and \textit{WISE} space telescopes \citep[see, e.g.,][KL14 hereafter]{Allen2004,Gutermuth2008,Esplin2014,Koenig2014}. However, the evolutionary stages associated with mid-IR excesses represent a critical phase, but not the entirety, of a star's pre-main sequence evolution.  IR excesses typically disappear before the star reaches the zero-age main sequence, as the circumstellar material accumulates into larger bodies or is dispersed by radiative or dynamical effects. A typical time scale for disk dispersal is $10^6$ years for stellar masses up to $3 M_{\odot}$. For masses larger than $7 M_{\odot}$, the time scale drops by an order of magnitude \citep{Gorti2009}. Censuses of YSOs that rely exclusively on the identification of IR excess sources will, therefore, be substantially incomplete for stellar populations as young as $\sim$3-5 Myrs of age \citep[e.g.][]{Haisch2001}. In addition to the explicit evolutionary bias of an IR-selected cluster census, the influence of neighboring stars on the dispersal of circumstellar material via dynamical interactions and/or radiative effects \citep[e.g.,][]{Pfalzner2014} could introduce complex secondary spatial biases in a purely IR-selected cluster census. 

Optical variability provides a complimentary means to identify pre-main sequence stars. Optically visible YSOs with substantial amounts of circumstellar material exhibit optical and near-IR variations with the largest characteristic amplitudes, driven by variations in optical emission and dust heating from accretion activity, or absorption and obscuration by circumstellar dust \citep[e.g.][]{Grankin2007, Grankin2008, Oliveira2008, Scholz2009, Rice2015, Rodriguez2015}.  Optical variability persists at lower amplitudes, however, even after circumstellar material has dissipated \citep[e.g.,][]{Grankin2008, Xiao2012}. Color information indicates these variations are driven by the presence of non-axisymmetric distributions of starspots on the photospheres of these young, magnetically active stars, which produce photometric variations as the star rotates and the spot groups transit the visible portion of the stellar disk \citep[e.g., ][]{Covey2016}. This spot-induced photometric variability has proven to be an effective means of identifying pre-main sequence stars \citep{Briceno2005}.  

In this paper, we describe the observational design and targeting methods utilized in the APOGEE-2 survey of the Orion Star Forming Complex. In this effort, we have constructed and validated a sample of candidate YSOs selected from optical and infrared wide-field photometric surveys covering the entirety of the OSFC, providing a consistent sample for studying spatial gradients in stellar kinematics, ages, and compositions. These uniformly selected YSOs received our highest priority, but were supplemented with additional YSOs identified in prior surveys of sub-regions of the OSFC. The original IN-SYNC survey of Orion A relied primarily on legacy catalogs for target selection and did not include a uniformly selected sample.  As a result, while IN-SYNC observations exist for many of the uniformly selected targets in Orion A, new observations of the cloud have been incorporated into the APOGEE-2 program to ensure a fully consistent approach to targeting and observing this critical subset of our survey sample. Subsequent papers in this series will combine radial velocities and stellar parameters measured from these APOGEE spectra with \textit{Gaia} photometry and astrometry to analyze the spatial distribution, kinematics, and star formation histories of the YSOs and stellar populations that make up the OSFC. Our first efforts, which are now underway, use this homogeneous dataset to identify coherent spatial and kinematic substructures across the full OSFC (Kounkel et al., in prep); subsequent analyses will provide finer-grained diagnoses of the membership and properties of each sub-population. 

This paper is structured as follows:  In Section \ref{sec:observations} we describe the observational data used to generate our catalog of uniformly selected candidate YSOs; we also summarize the APOGEE-2 observations that were obtained in early 2016, which together with the legacy IN-SYNC Orion A observations provide a first indication of the yield of our targeting methods. In Section \ref{sec:uniform} we describe the construction of the catalog of uniformly selected YSOs, presenting the selection of IR excess and optically variable YSO candidates in Sec. \ref{sec:KL14explain} and \ref{sec:PanSTARRS} respectively. Section \ref{sec:supplementalSources} summarizes the additional OSFC members/candidates that were targeted to increase the completeness of our final target lists, albeit with a significantly more complex selection function, when fibers remain to be filled in a given field.  The process for producing final target lists in each field is presented in Section \ref{sec:finalPlates}. In Section \ref{sec:RVvalid} we quantify the yield of our uniformly selected targets by analyzing radial velocities measured in the IN-SYNC survey of Orion A \citep{DaRio2016, DaRio2017} and in the 2015-2016 APOGEE-2 observations of Orion B, Orion OB1a/b and Lambda Ori.  Finally, we summarize our conclusions and future analysis plans in Section \ref{sec:conclusions}. 

\section{Observations} \label{sec:observations}

\subsection{Data from the Literature: WISE+2MASS point sources}

Near- and mid-infrared (NIR, MIR) photometry used to select APOGEE-2 YSO targets was primarily drawn from the AllWISE Source Catalog \citep{Cutri2013}.  The AllWISE Source Catalog is a primary data product of the \textit{Wide-field Infrared Survey Explorer} \citep[WISE]{Wright2010}. The AllWISE data release combines data from the WISE cryogenic \citep{Wright2010} and post-cryo NEOWISE \citep{Mainzer2011} missions; the resulting WISE detections were associated with near-infrared counterparts in the 2MASS Point and Extended Source Catalogs \citep{Skrutskie2006} using a 3 arcsecond matching radius. Sources detected in both catalogs provide reliable photometry in up to seven wavelength bands spanning 1-22 $\mu$m, as required for the algorithm developed by KL14 to select candidate YSOs with infrared excesses (Sec. \ref{sec:KL14explain}).  

Additional YSO candidates were targeted for APOGEE-2 observations based on criteria that did not require a MIR detection (e.g., optical variability or location in an optical/near-infrared color-magnitude diagram). NIR photometry was still essential, however, to plan and prioritize the H-band APOGEE observations. For this purpose, we used a 3 arcsecond matching radius to identify counterparts to optical sources in the 2MASS point source catalog, which is effectively deeper for non-IR-excess-bearing pre-main sequence stars than the mid-IR flux-limited AllWISE catalog. 

\subsection{Optical photometry: Pan-STARRS1 3$\pi$ Survey}

We identified optically variable candidate YSOs throughout the OSFC using multi-epoch optical photometry from the Pan-STARRS1 (PS1) 3$\pi$ survey. The Pan-STARRS1 3$\pi$ survey was carried out using the GPC1 camera \citep{Hodapp2004,Onaka2008,Tonry2009} mounted on a 1.8m telescope at Haleakala, Hawaii. The camera provides images of a 3$^{\circ}$ field of view in five broad filters: $g_{\textrm{P1}}$,  $r_{\textrm{P1}}$,  $i_{\textrm{P1}}$,  $z_{\textrm{P1}}$ and $y_{\textrm{P1}}$, with effective wavelengths of 4800\AA\, 6200\AA\,, 7500\AA\,, 8700\AA\, and 9600\AA\ respectively. Source photometry is measured by the Pan-STARRS1 Image Processing Pipeline \citep{Magnier2006, Magnier2007, Magnier2008, Magnier2013}, which provides relative and absolute photometry accurate to better than 1\% \citep{Tonry2012, Schlafly2012}. The photometry used in this analysis were drawn from the third processing version release (PV3). 

\subsection{APOGEE Spectroscopy \label{AP2N}}
Observations for the APOGEE-2 Young Cluster Survey are collected with the APOGEE-2 northern spectrograph \citep[APOGEE-2N hereafter; ][]{Wilson2010} on the 2.5m SDSS telescope at the Apache Point Observatory. The APOGEE-2N spectrograph enables the simultaneous acquisition (via optical fibers plugged into pre-drilled aluminium plug-plates) of up to 300 moderate-to-high resolution (R $\sim$ 22,000) H-band (1.51-1.7 $\mu$m) spectra across a 1.5 deg. radius field-of-view. Spectra are processed by the APOGEE data reduction \citep{Nidever2015} and APOGEE Stellar Parameter and Chemical Abundances \citep[ASPCAP;][]{GarciaPerez2016} 
pipelines, to produce basic APOGEE-2 data products, including calibrated spectra, stellar parameters (e.g., $T_{eff}$, log~$g$), radial velocities, and bulk and individual elemental abundances \citep{Holtzman2015}.  Analysis routines developed by \citet{Cottaar2014} can perform an independent spectral analysis to extract joint constraints for a suite of spectral parameters most important for young stars: effective temperatures, surface gravities, and radial velocities, as well as projected rotational velocities ($v$ sin $i$), $H$-band veiling flux (r$_H$), and by combining the APOGEE-based stellar parameters with 2MASS photometric measurements, infrared color-excesses ($E(J-H)$). 

%% ----- APOGEE field table -----
\startlongtable
\begin{deluxetable}{lcccc}
\tablewidth{\columnwidth}
\tabletypesize{\scriptsize}
\tablecaption{Fields in APOGEE-2 Orion Survey\label{tab:APOGEE-2_OrionFields}}
\tablehead{ 
\colhead{Field} & \colhead{R.A.} & \colhead{Dec.}  & \colhead{Plate} & \colhead{Epoch} \\
\colhead{Name} & \colhead{(deg.)} & \colhead{(deg.)}  & \colhead{ID} & \colhead{(MJD)} 
}
\startdata
$\lambda$ Ori A & 84.140 & 10.340 & 8879 & 2457406 \\
 &  &  & \nodata & 2457411 \\
&  &  & 8880 & 2457413 \\
&  &  & \nodata & 2457648 \\
&  &  & \nodata & 2457649 \\
&  &  & 8881 & 2457650 \\
&  &  & 8882 & 2457651 \\
&  &  & 8882 & 2457655 \\
&  &  & 8883 & 2457656 \\
&  &  & 8883 & 2457675 \\
&  &  & 8884 & 2457676 \\
&  &  & 8884 & 2457677 \\
\hline 
$\lambda$ Ori B & 82.340 & 11.730 & 8885 & 2457409 \\
 &  &  & 8886 & 2457411 \\
 &  &  & \nodata & 2457413 \\
&  &  & 8887 & 2457678  \\
\hline 
$\lambda$ Ori C & 86.611 & 8.993 & 9482 & 2457685 \\
\hline 
Orion B-A & 86.654 & 0.134 & 8890 & 2457433 \\
 &  &  & 8891 & 2457443 \\
 &  &  & 8892 & 2457675 \\
 &  &  & \nodata & 2457677 \\
 &  &  & 8893 & 2457677  \\
 &  &  & \nodata & 2457678  \\ 
\hline 
Orion B-B & 85.416 & -2.120 & 8894 & 2457433 \\
 &  &  & 8895 & 2457434 \\
 &  &  & 8896 & 2457435 \\
 &  &  & 8897 & 2457436 \\
 &  &  & 8898 & 2457654 \\
 &  &  & 8899 & 2457680 \\
\hline 
Ori OB1ab-A & 84.099 & -2.201 & 9468 & 2457794 \\
 &  &  & 9468 & 2457795 \\
\hline 
Ori OB1ab-B & 83.997 & 0.695 & 9471 & 2457683 \\
 &  &  & 9472 & 2457684 \\
 &  &  & 9473 & 2457707 \\
 &  &  & 9474 & 2457708 \\
 \hline 
Ori OB1ab-C & 82.502 & -1.501 & 9475 & 2457732 \\
 &  &  & \nodata & 2457734 \\
 &  &  & 9476 & 2457764 \\
 &  &  & 9477 & 2457796 \\
 \hline 
Ori OB1ab-D & 80.697 & -1.799 & 9478 & 2457708 \\
 &  &  & 9479 & 2457713 \\
\hline 
Ori OB1ab-E & 81.496 & 1.006 & 8900 & 2457648 \\
 &  &  & 8901 & 2457649 \\
 &  &  & 8901 & 2457650 \\
 &  &  & 8902 & 2457652 \\
 &  &  & 8902 & 2457653 \\
 &  &  & 8903 & 2457675 \\
\hline 
Ori OB1ab-F & 82.000 & 3.000 & 8904 & 2457676 \\
 &  &  & 8905 & 2457410 \\
 &  &  & 8906 & 2457411 \\
 &  &  & \nodata & 2457412 \\
\hline
Orion A-A & 84.100 & -5.100 & \nodata & \nodata \\
\hline
Orion A-B & 83.550 & -5.300 & 9533 & 2457737 \\
&  &  & \nodata & 2457762 \\
&  &  & \nodata & 2457794 \\
\hline
Orion A-C & 84.251 & -6.899 & 9659 & 2457790 \\
&  &  & \nodata & 2457792 \\
&  &  & \nodata & 2457793 \\
\hline
Orion A-D & 84.500 & -7.200 & 9660 & 2457790 \\
&  &  & \nodata & 2457792 \\
&  &  & \nodata & 2457793 \\
\hline
Orion A-E & 85.200 & -8.700 & 9661 & 2457795 \\ 
\enddata

\end{deluxetable}

\begin{figure}[t!] 
\centerline{\includegraphics[angle=0,width=3.5in]{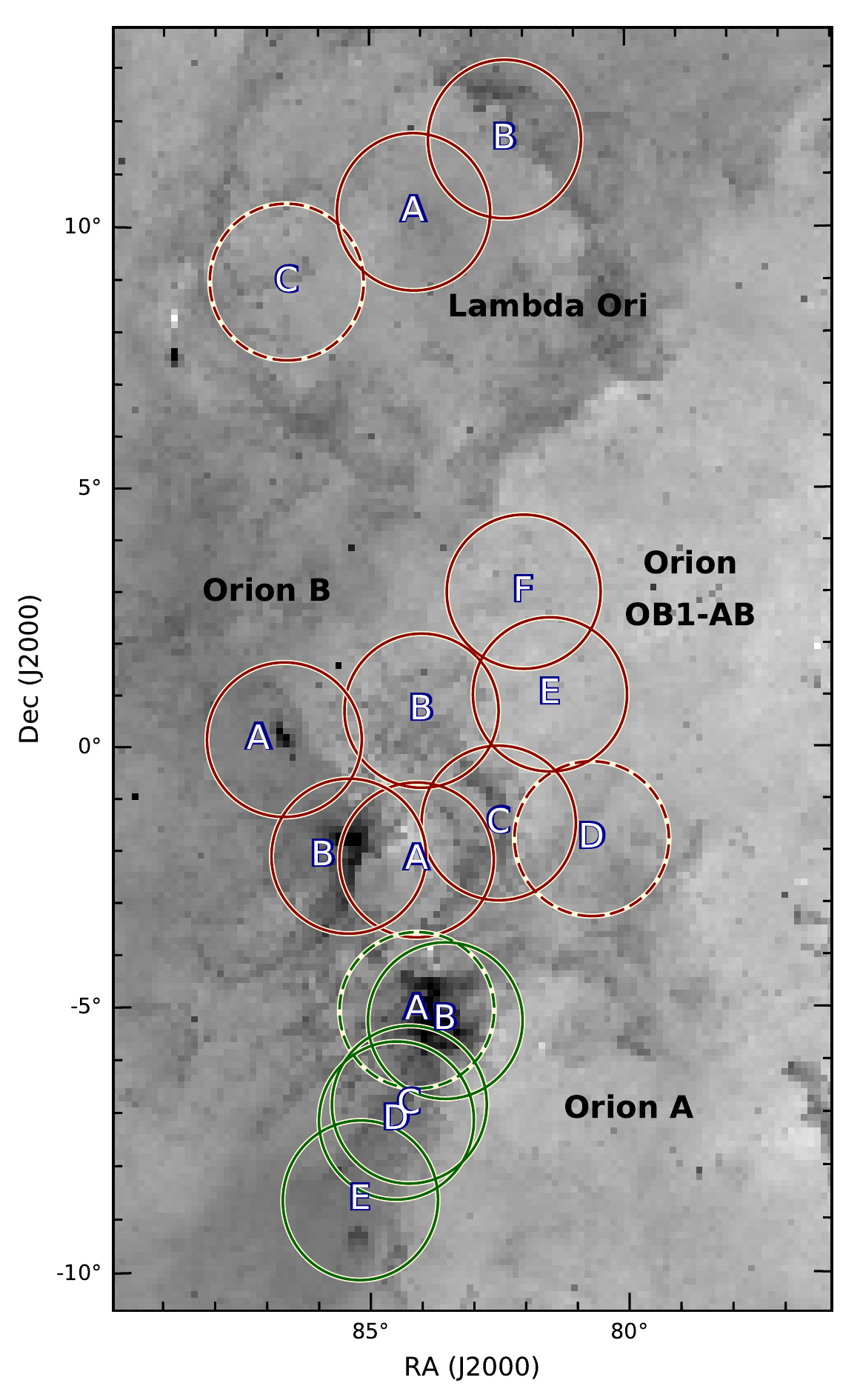}}  

\caption{Field plan for APOGEE-2 Young Cluster Survey in Orion. OSFC main regions are indicated with bold labels. In each subregion, individual fields are identified by letters. Plates indicated with a dashed line are only partially observed at the time of publication and are expected to be completed in Winter 2017-18. Fields in Orion A (shown as green circles) were first observed in APOGEE-1, as reported by \citet{DaRio2016}, with additional observations planned for APOGEE-2. The background image is a mosaic from the WISE 12$\mu m$ WSSA survey \citep{wssa}}
\label{fig:fieldplanHighligh}
\end{figure}

The OSFC is the largest region targeted in the APOGEE-2 Young Cluster Survey. The complete 16 field plan for the APOGEE-2 Orion Survey is outlined in Table \ref{tab:APOGEE-2_OrionFields} and shown schematically in Figure \ref{fig:fieldplanHighligh}. The first APOGEE-2 observations in Orion were conducted in January and February of 2016, and comprise fifteen 1-hour `visits'\footnote{A `visit' is a contiguous sequence of short exposures on the same plate, and represents the fundamental unit of the APOGEE observing schedule and survey plan; see \citet{Zasowski2013} for more details.} for twelve distinct plate designs in five of our planned fields, sampling three sub-populations within Orion: $\lambda$ Ori, Orion B, and the  Orion OB1a/b region (covering $\sigma$-Ori and 25 Ori). Additional observations, comprising the bulk of the planned surveys for the $\lambda$ Ori, Orion B, and Orion OB1ab fields, were completed during the 2016-2017 observing season. We list the MJDs of all observations acquired for APOGEE-2 Orion fields through the summer of 2017 in Table 1. Other star forming regions and young clusters targeted by this APOGEE-2 Goal Science program are Taurus, NGC 2264, Alpha Per, and the Pleiades; additional regions may be able to be added to the program in 2018-2019, depending on survey progress.

APOGEE observations in crowded environments face an important trade-off between sample completeness and typical S/N. The width of APOGEE-2N fibers and their plugging ferrules prevent observations of two sources separated by less than 72\arcsec~on a single APOGEE-2N plate. A more complete sample of crowded cluster stars can be achieved by designing distinct plates for each visit, so that objects whose fibers would collide on a single plate design can be observed on separate plates. Allocating a fiber on only one plate design, however, limits the S/N that can be achieved for a source of a given magnitude.  In particular, a typical 1-hour plate visit returns S/N $\geq$ 100 for sources with $H<$ 11 mag, whereas a S/N $\sim$ 100 spectrum for an $H\sim$ 12 mag source must be built up by co-adding three separate 1-hour (and S/N $\sim$ 45) spectra. We choose to limit our sample to sources with $ H < 12.4$ mag to ensure a six visit source will return S/N $\sim$100 \citep[see Fig. 22 by][for APOGEE's S/N performance as a function of magnitude for 6 visit sources]{Nidever2015}. For more information on the standard APOGEE plate design and observing strategies, see \citet{Zasowski2013}.

To reduce the limitation imposed by fiber crowding and maximize the number of bona-fide OFSC members observed in this program, we designed several distinct plates for each of the fields in the Orion Complex. To achieve a more uniform S/N in the final combined spectra of sources of different brightnesses, we identified the minimum number of visits we sought to achieve for each source as a function of its H-band magnitude: one visit suffices for $H <$ 11 mag sources, whereas three visits is the goal for 11 $\leq$ $H <$ 12.2 mag sources, and six visits is the goal for sources with $H \geq$ 12.2 mag. As described in Section \ref{sec:finalPlates}, targets were prioritized according to the criteria used to include them in the survey sample. Fibers were assigned to sources according to their prioritization, allocating fibers on distinct visit plate designs up to the number specified as their fiber goal. Efforts were made to avoid collisions with ancillary science targets \citep[e.g., background stars targeted to continue the survey of the Galactic extinction law begun by][]{Schlafly2016}; a negligible number of targets may also be lost due to fiber collisions with calibrators on the final plates designed by the APOGEE-2N targeting team. The final plates designed for each visit to each field included $\leq$260 primary science targets, leaving $\geq$40 fibers available for telluric and sky subtraction purposes.

\begin{figure}[t!]
\centerline{\includegraphics[angle=0,width=4.5in]{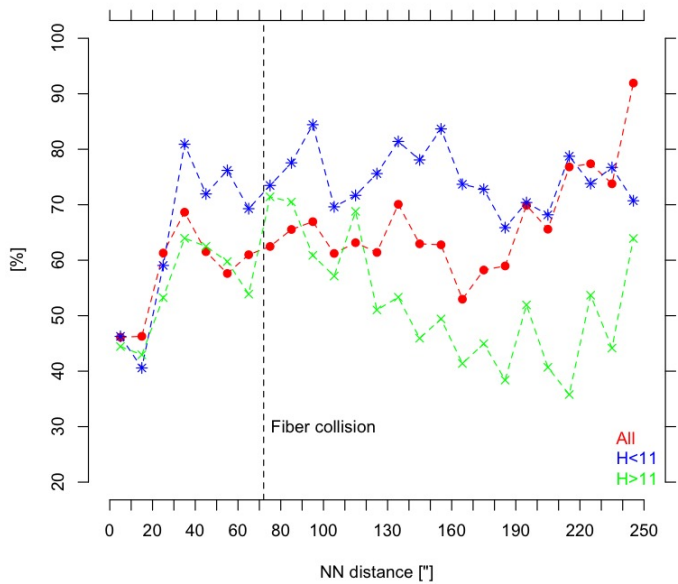}}
\caption{Ratio between the nearest neighbor distance distributions for all candidates in the observed fields and those targets with at least one visit. Symbols show the completeness for different subsets of the sample, including the complete sample (red dots), bright stars ($H<11$; blue asterisks), and faint stars ($H>11$; green crosses). The mean completeness of each sample falls at $Dnn<35^{\prime\prime}$}.
\label{fig:Dnn}
\end{figure}

To quantify spatial biases in our survey due to APOGEE-2N's $72^{\prime\prime}$ fiber collision limit, we performed a simple nearest neighbor analysis on our final targeted sample. We computed the nearest neighbor distances ($Dnn$) for the complete sample of candidates in the observed fields (i.e., including all candidates prior to the the final fiber assignment stage) and for the final sample of targets which were targeted for observation with at least one visit. We then computed the ratio of the two distributions as a function of nearest neighbor distance to quantify the completeness of the target sample. The results, which are shown in Figure \ref{fig:Dnn}, indicate that the sample is indeed less complete for smaller nearest neighbor separations.  Our strategy to use multiple independent observations of a field has mitigated the effect, however, such that the completeness does not drop until nearest neighbor distances significantly below the 72$^{\prime\prime}$ fiber collision limit: the effect is most clearly seen at a typical separation of $\sim$35$^{\prime\prime}$, where the completeness declines from $\sim60\%$ for $Dnn > 35^{\prime\prime}$ to $\sim40\%$ for $Dnn < 35^{\prime\prime}$.  Our completeness is even higher for uniformly selected YSOs, which received the highest priority for fiber assignment; even in the dense Orion A + B fields, we achieved completeness of $\gtrsim$ 85\% for $Dnn > 25^{\prime\prime}$ and $\gtrsim$ 60\% for $Dnn < 25^{\prime\prime}$.  

As the number of visits required to meet our signal-to-noise goal is a direct function of source magnitude, we also performed this nearest neighbor analysis for subsets of bright ($H<11$) and faint ($H>11$) targets. The completeness levels achieved for these subsets of our sample are also shown in Figure \ref{fig:Dnn}, and indicate that the sample of bright stars is $\sim$10-15\% more complete than the sample of faint stars, for all separations. 

In summary, the target sample is moderately biased against stars with the closest neighbors. In both the full and uniformly selected samples, the completeness drops by $\sim20\%$ for sources with nearest neighbor separations less than $35^{\prime\prime}$ and $25^{\prime\prime}$, respectively. We emphasize that our scientific goals do not require a \textit{complete} survey, but rather an \textit{homogeneously selected} sample of young stars. Nevertheless, our sample should include 40-60\% of spatially resolved wide binaries with separations as small as ~2000 AU ($5^{\prime\prime}$ resolution). Even in the densest regions such as Orion A  and B, our targets include more than half of the stars and the sample is representative enough to describe well the properties of the full stellar population.

\begin{figure*}[t!] 
\centerline{\includegraphics[angle=0,width=6in]{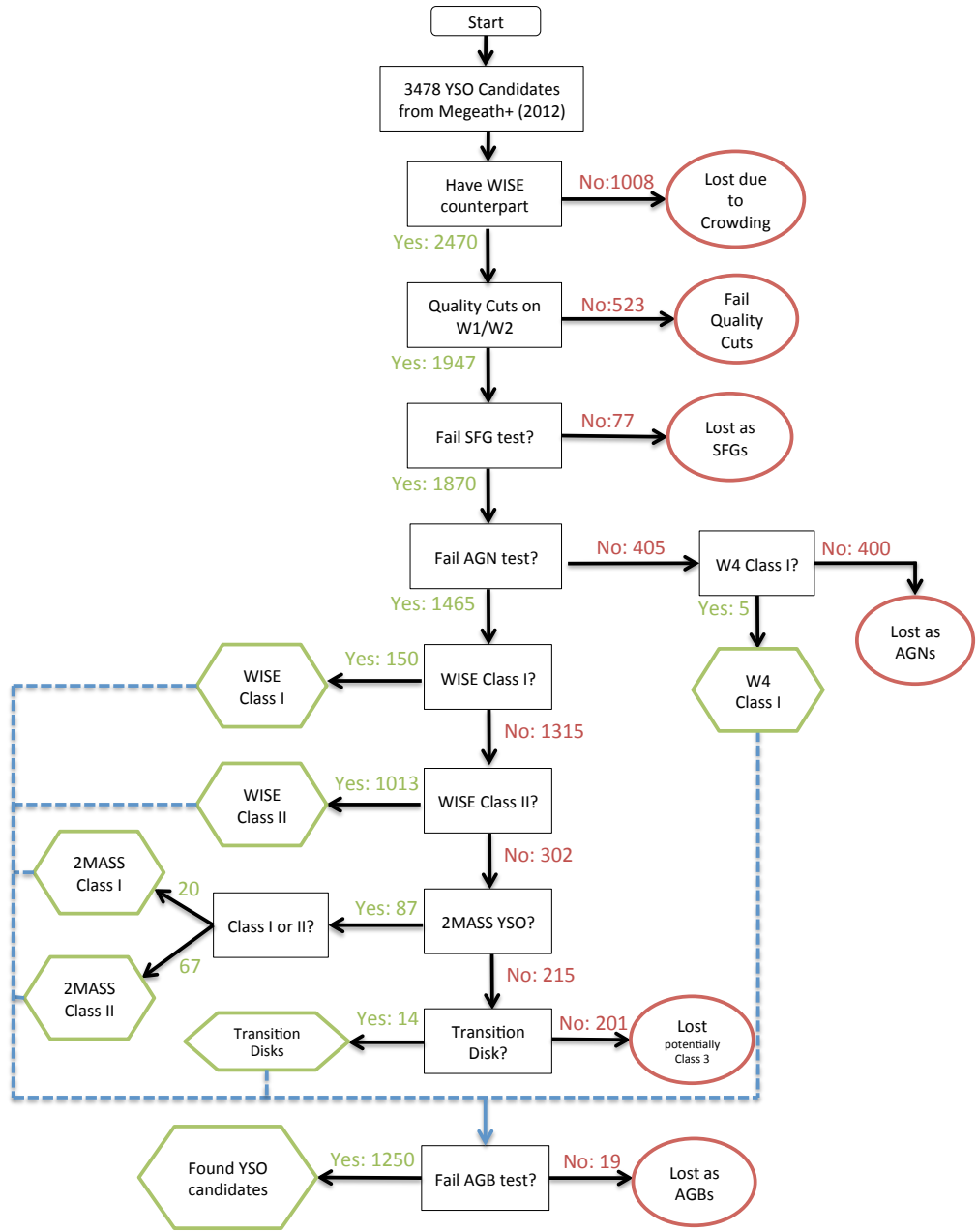}}
\caption{Summary flowchart of the KL14 algorithm (see description in Sec. \ref{sec:KL14explain}) as applied to the catalog of IR-excess sources compiled by \citet{Megeath2012} from Spitzer photometry of Orion A \& B. Sources in the \citet{Megeath2012} catalog that are not recovered by the KL14 classification are primarily lost either because they lack a WISE counterpart, have low quality measurements in W1 and W2, or are classified by the KL14 system as likely AGN. The number of sources from the \citet{Megeath2012} catalog that survive or are discarded at each step of the KL14 classification is shown in the flowchart with the `Yes:' and `No:' labels, respectively. Of the 3478 YSOs in the \citet{Megeath2012} catalog, 1250 are classified as candidate YSOs by the KL14 algorithm and 2228 either lacked the required data or received a non-YSO classification. }
\label{fig:megeathFlowchart}
\end{figure*}

\section{Methods for Uniformly Selecting Young Stellar Objects}\label{sec:uniform}

To select targets in each APOGEE-2 OSFC field, we adopted a multi-tier prioritization scheme.  The highest priority targets were YSO candidates selected in a consistent manner across the entire survey footprint using data from wide-field surveys. In this section we describe the construction of the catalog of `uniformly selected' candidate YSOs that provide these highest priority targets; in future sections we describe how these uniformly selected targets are supplemented with previously confirmed members, and new candidates selected via (lower yield) optical+near-infrared color-magnitude criteria.

\subsection{IR Excess: The Koenig \& Leisawitz 2MASS+WISE Algorithm} \label{sec:KL14explain}

KL14 developed a method to select YSOs with evidence of circumstellar disks revealed in all-sky 2MASS and WISE photometry.  \citet{Koenig2015} then spectroscopically validated the fidelity of this selection method, finding that $\sim$80\% of the candidate YSOs in $\sigma$ and $\lambda$ Ori selected by this method are indeed probable or likely members of the OSFC. In this section, we briefly summarize the selection method developed by KL14, which we outline schematically in Figure \ref{fig:megeathFlowchart}. We then validate the algorithm's yield with training sets of known YSOs in Taurus and Orion, before performing a blind search for sources across the full Orion complex.  

\begin{figure}[t!] 
\centerline{\includegraphics[angle=0,width=\columnwidth]{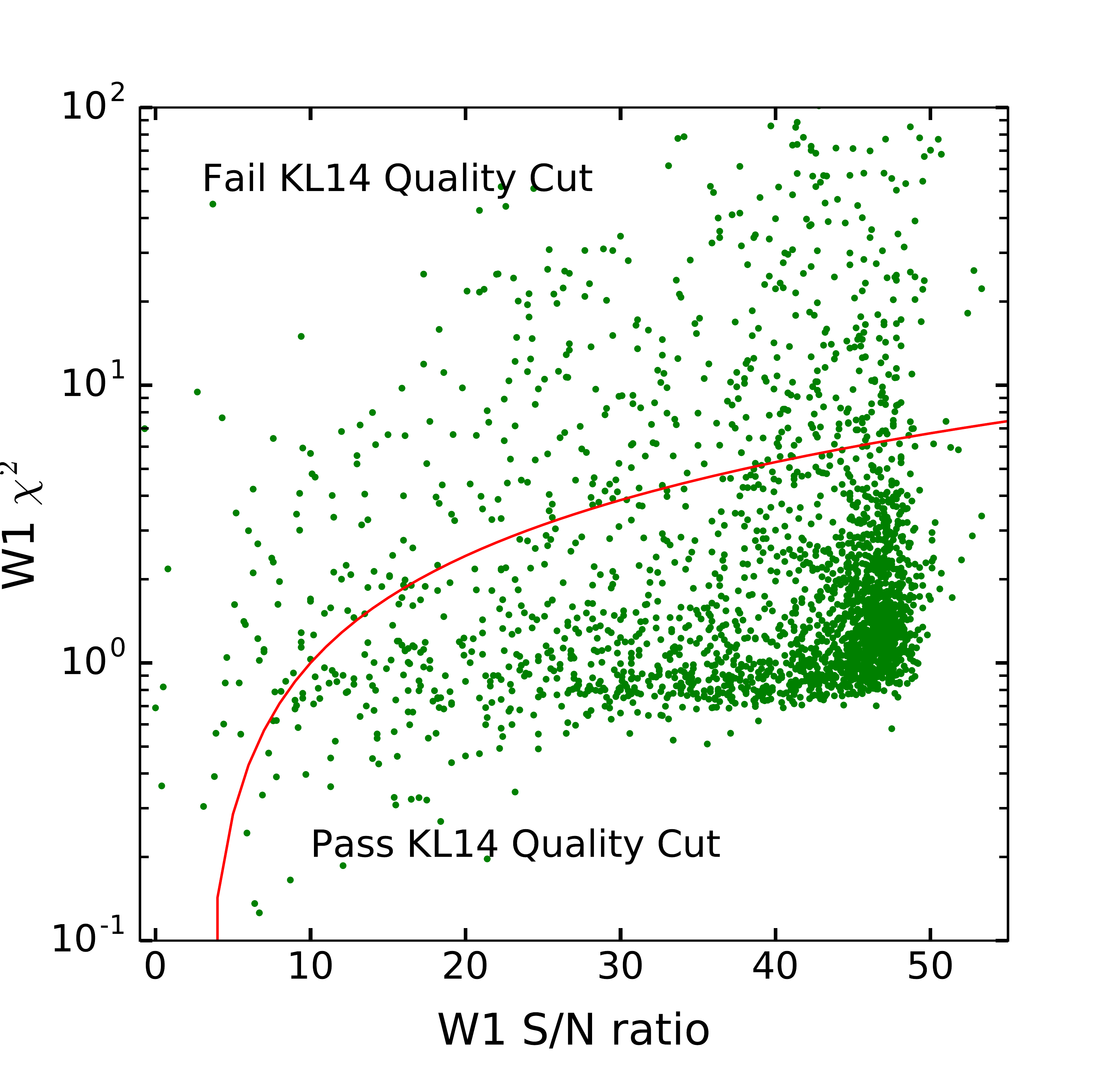}}
\caption{W1 S/N ratio (W1$_{snr}$) vs. W1 reduced $\chi^2$ value (W1 $\chi^2$) for 2470 WISE counterparts, with the KL14 quality cut ($w1\chi^2 = (W1_{snr} - 3)/7$) overlaid as a red line. More than 21\% (523/2470) of the WISE sources fail the quality cut and are eliminated from further classification.  This quality cut reduces contamination at a cost of reduced completeness in active star forming regions.}
\label{fig:qualitycuts}
\end{figure}

The first several steps in the KL14 classification scheme serve to eliminate low quality photometric detections, or sources that appear consistent with a non-YSO nature. The KL14 algorithm is designed to assess sources that are detected in both WISE and 2MASS; in practice, the depth of the 3-band (W1, W2, W3) WISE detections is the primary factor limiting the size of the sample that can be classified using the KL14 technique. For sources with WISE counterparts, the KL14 algorithm first applies a series of photometric quality cuts to reject sources with low quality detections in W1 or W2 (i.e., null measurement uncertainty in W1 [w1sigmpro $=$ null] or W2 [w2sigmpro $=$ null]) or a high reduced $\chi^2$ given the source's signal-to-noise ratio in that band.  Figure \ref{fig:qualitycuts} provides an example of applying this latter cut to W1 detections, where sources are rejected if w1$\chi^2 > \frac{W1_{snr} -3}{7}$. Similar cuts are applied to W2, W3, and W4 photometry; see Sec. 3.1.1 in KL14 for the relevant equations. After rejecting potentially spurious WISE detections, the KL14 algorithm rejects sources identified by color-magnitude cuts in the (W1-W2) vs. (W2 - W3) and (W1-W3) vs W1 space as likely Star Forming Galaxies (SFGs) and Active Galactic Nuclei (AGN), respectively. 

\begin{figure}[t!] 
\centerline{\includegraphics[angle=0,width=\columnwidth]{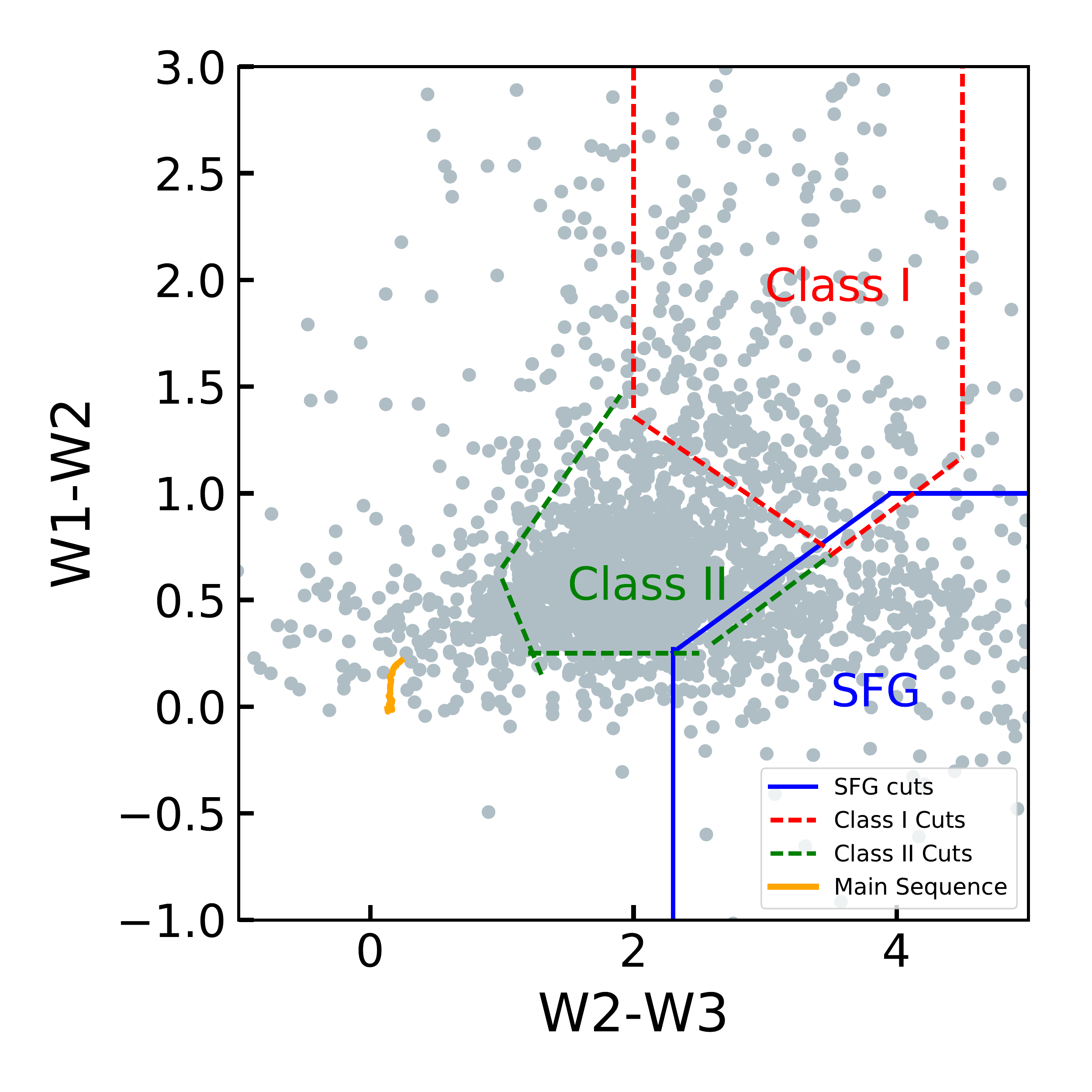}}
\caption{(W1-W2) vs. (W2-W3) color-color diagram, populated by candidate YSOs identified by \citet{Megeath2012} that meet the KL14 WISE photometry quality cuts shown in Figure \ref{fig:qualitycuts}. Colored lines indicate regions used by KL14 to classify Class I (red lines) and Class II (green lines) YSOs, as well as Star Forming Galaxies (SFGs; blue lines). Of these 1140 sources, 77 objects with W1 $>$ 13 are classified as SFGs and removed by the KL14 algorithm prior to classification. Of the remaining sources, 150 are classified as Class I and 1013 as Class II. The KL14 selection selects Class I and Class II objects with high fidelity. WISE colors of typical main sequence sources, as tabulated by \citet{Davenport2014}, are shown as a yellow locus. }
\label{fig:megeathClassification}
\end{figure}

After eliminating low-quality or likely contaminant sources, the next steps in the KL14 classification scheme use a series of NIR/MIR color-color cuts to identify likely YSOs in a range of evolutionary states.  Sources whose WISE or 2MASS colors are consistent with a Class I and Class II classification (based on cuts in the $H - K_S$ vs (W1 - W2) and the (W1 - W2) vs (W2 - W3) color-color spaces; see Figure \ref{fig:megeathClassification} for an example of the latter cut) are provided a provisional YSO classification. The remaining sources, whose W1-W3 Spectral Energy Distribution (SED) did not indicate the presence of an IR excess, are then flagged as provisional transition disk candidates and added to the provisional YSO list if they satisfy the requirements as given in KL14 Sec. 4.2.2. A final set of possible Class I sources are extracted from the likely AGN sample with color cuts on (W1 - W2), (W2 - W4), and (W3 - W4) colors. These sources are added to a provisional YSO list, which is then subjected to a final screening that uses the W1 vs (W1 - W2) color-magnitude and (W1 - W2) vs (W3 - W4) color-color diagrams to identify and remove any remaining Asymptotic Giant Branch (AGB) candidates.  Sources that pass all the above tests are ultimately identified as likely YSOs, and given a likely Class designation based on their mid-IR SED slopes.

\begin{figure*}[t!] 
\centerline{\includegraphics[angle=0, width = 6in]{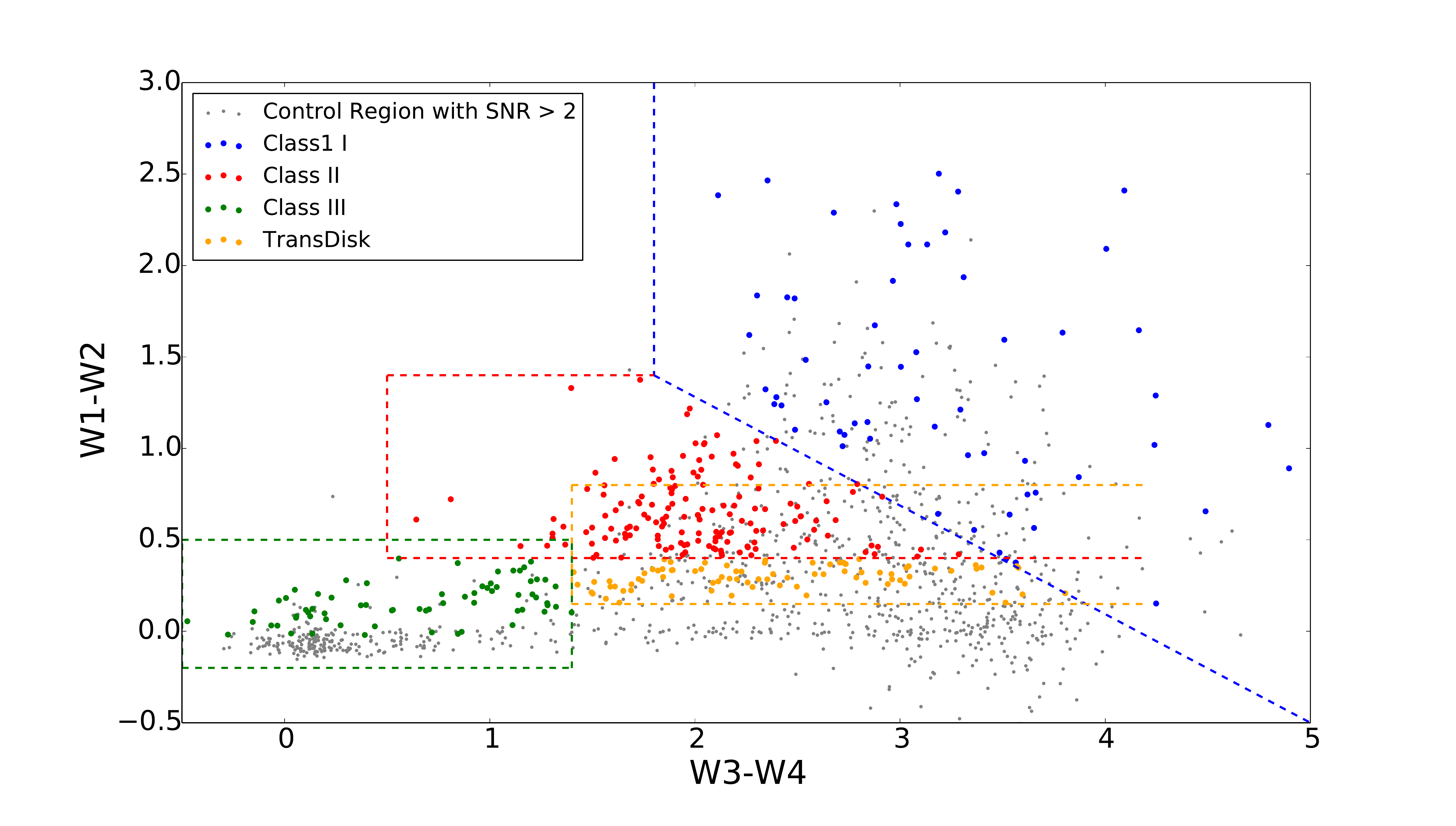}}
\caption{(W1-W2) vs. (W3-W4) color-color diagram for all sources in the control region with SNR $> 2$ in all bands (greyscale points), overlaid with YSOs cataloged by \citet{Esplin2014} (colored points). The known YSOs are color coded with classifications from a set of simple color cuts shown as dashed lines. While many of these control sources would be identified as likely YSOs using simple color-cut criteria, the KL14 algorithm correctly eliminates all but 7 as non-YSOs. }
\label{fig:EsplinSimpleColor}
\end{figure*}

\subsubsection{Validation in Taurus}

\begin{figure}[t!] 
\centerline{\includegraphics[angle=0,width=\columnwidth]{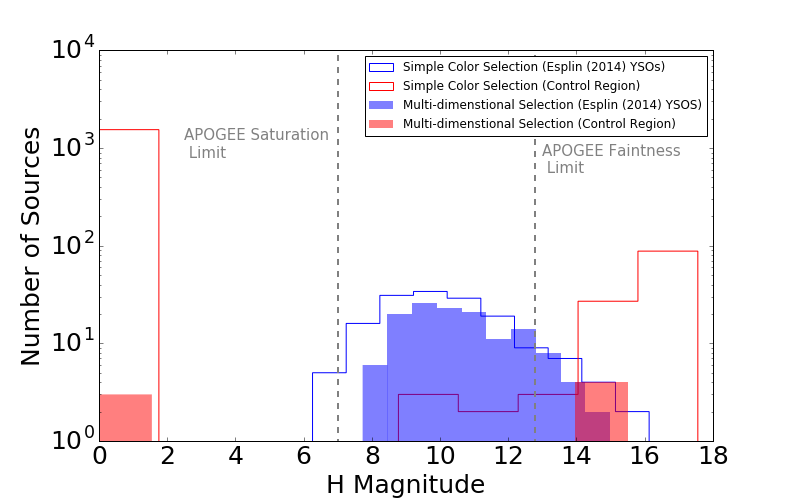}}
\caption{Histograms of $H$ magnitudes for known Class II YSOs in Taurus from \citet{Esplin2014}(blue histograms) and candidate Class II sources identified in the off-cloud control region (red histograms).  Open histograms show sources identified with the simple (W1-W2) vs. (W3-W4) color cuts shown in Figure \ref{fig:EsplinSimpleColor}; filled histograms show sources of each type identified as candidate YSOs by the KL14 selection algorithm. }
\label{fig:EsplinHHist2}
\end{figure}

As a first test of the completeness and fidelity of the KL14 selection method, we examined the algorithm's ability to recover YSOs with evidence of circumstellar disks in the Taurus star forming region.  2MASS and WISE counterparts were identified for 414 known Taurus members compiled by \citet{Esplin2014} using a 1\arcsec ~matching radius; sources lacking unique detections in both 2MASS and all four WISE bands were eliminated from further consideration. To evaluate the number of false positives that the KL14 algorithm produces in regions free of active star formation, we also analyzed a catalog of 2MASS-WISE detections in a 2 square degree off-cloud field (control region; Dec: 24 - 26 $\deg$; RA: 75 - 77.5 $\deg$). The WISE color cuts used to assign YSO classifications to sources in the \citet{Esplin2014} catalog and to identify potential contaminants in the off-cloud region are shown in Figure \ref{fig:EsplinSimpleColor}.  Of the 156 bona fide Taurus members in the Esplin catalog that satisfy the W1-W2 vs. W3-W4 criteria for Class II YSOs shown in Figure \ref{fig:EsplinSimpleColor}, 135 satisfy the criteria in the KL14 algorithm for identification as a YSO candidate; by contrast, only 7   (0.037\%) of the color-selected YSO candidates in the off-cloud region are retained as likely YSOs according to the KL14 criteria. As Figure \ref{fig:EsplinHHist2} shows, simple color-cut selection techniques flag a smaller, but non-trivial, number of candidate Class II YSOs in the off-cloud region than within Taurus itself; utilizing the more complex, multi-dimensional KL14 selection algorithm, however, preserves the vast majority of known Class II YSOs in Taurus while excluding nearly all candidate Class II sources in the off-cloud region. This test provides a first demonstration that the KL14 algorithm accurately retains a high fraction of known YSOs with infrared excesses, without spuriously flagging a large number of non-YSO interlopers or contaminants. 

\subsubsection{Validation in Orion}

\citet{Megeath2012} analyzed Spitzer/IRAC observations of the Orion A \& B clouds to compile a census of dusty YSOs throughout the high-extinction regions of the OSFC. The areas targeted by our APOGEE survey of the OSFC include the full extent of the \citet{Megeath2012} catalog, making it a valuable resource for testing the completeness of the KL14 selection method. We find our implementation of the KL14 selection, (a sample of which is shown in Table \ref{tab:IRmembership}) process identifies 44\% and 50\% of the Class I and II objects, respectively, cataloged by \citet{Megeath2012} and \citet{Fang2013} in Orion's high extinction regions. 

\begin{figure}[t!] 
\centerline{\includegraphics[angle=0,width=\columnwidth]{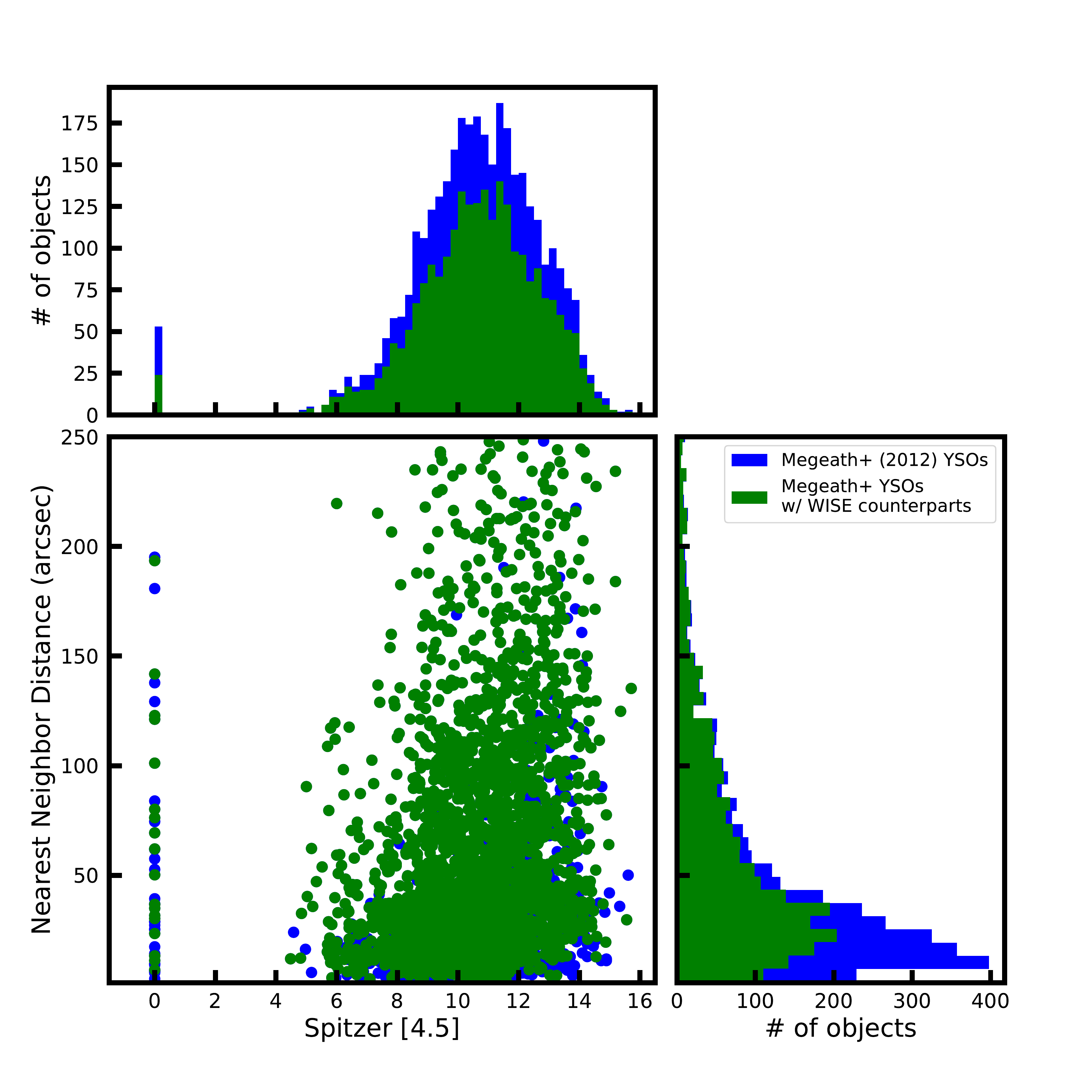}}
\caption{Nearest neighbor distance as a function of Spitzer [4.5] magnitude for sources identified as YSOs by \citet{Megeath2012}, distinguishing between all YSOs in the catalog (blue) and the subset with unique WISE counterparts (green points). Sources without a Spitzer [4.5] detection are shown at [4.5] = 0. The histograms in the top and right sub-panels show the number of YSOs in the \citet{Megeath2012} catalog with and without WISE counterparts as functions of each axis. The ratio of WISE counterparts to Megeath sources remains relatively constant for Spitzer [4.5] magnitude (top sub-panel) which suggests that incompleteness near or below WISE's flux limit is not responsible for the missing WISE counterparts for some Megeath sources. The ratio steadily decreases, however, for sources with smaller nearest neighbor distances (right sub-panel), reaching a minimum of about 0.5 at nearest neighbor distances close to WISE's angular resolution.  }
\label{fig:neighbor2}
\end{figure}

\begin{figure}[t!] 
\centerline{\includegraphics[angle=0,width=\columnwidth]{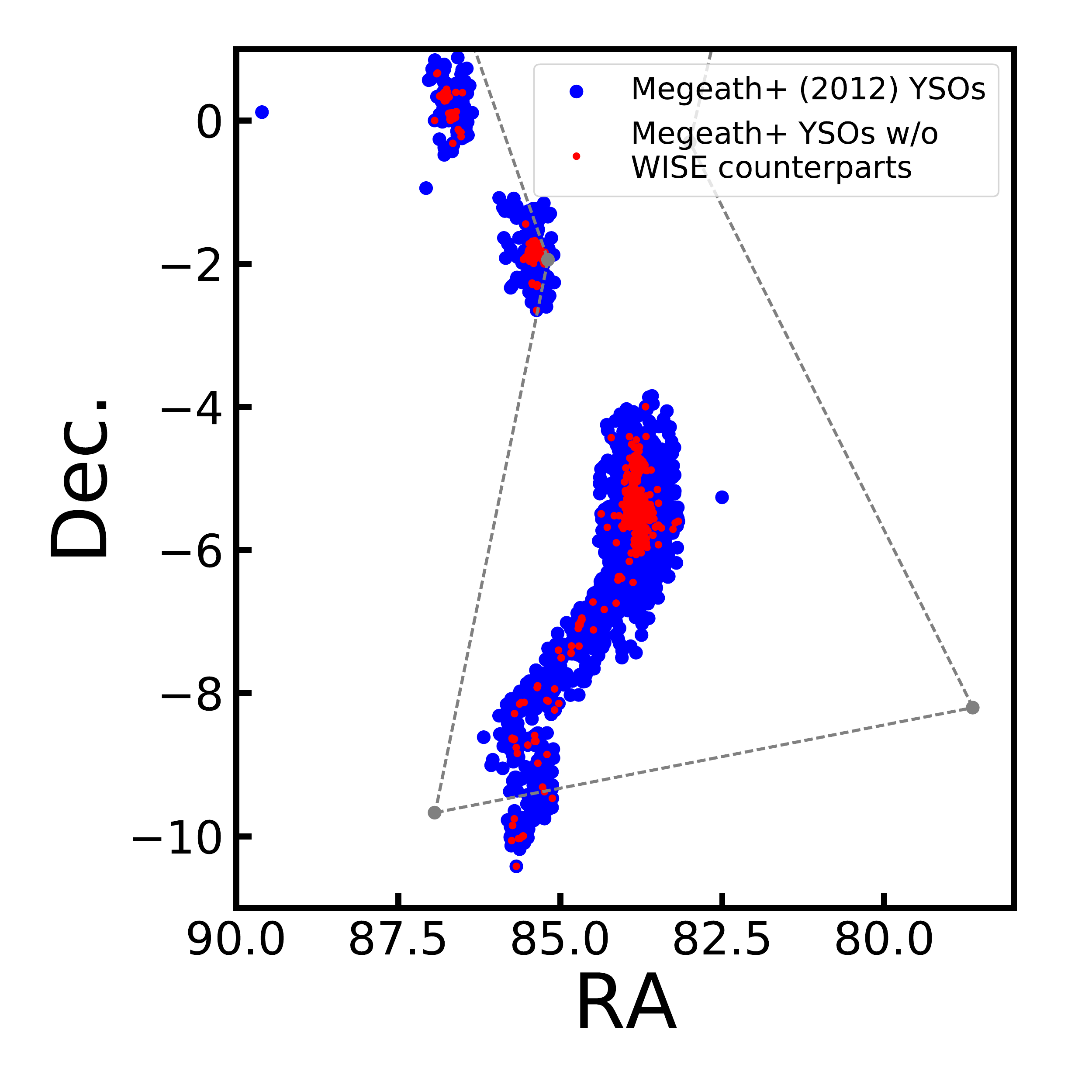}}
\caption{Locations of YSOs in Orion identified by \citet{Megeath2012}, split into sets of sources that do (blue points) and do not (red points) possess WISE counterparts. 
Most sources lacking counterparts are in regions with high number densities of YSOs leading to substantial crowding.}
\label{fig:noCounterpartsSpatial}
\end{figure}

The most significant factors limiting the KL14 algorithm's recovery of YSOs identified by \citet{Megeath2012} are the combined effects of crowding and WISE's low angular resolution ($\sim 6.5\arcsec$~in W3; $\sim12\arcsec$ ~in W4). As seen in Table \ref{tab:megeathResults}, 1008 sources (28\% of the \citet{Megeath2012} catalog) lack a unique WISE counterpart, which prevents their recovery with the KL14 algorithm. As Figure \ref{fig:neighbor2} demonstrates, the frequency of WISE counterparts to \citet{Megeath2012} sources drops for smaller nearest neighbor distances. Searching the 2MASS catalog, we find that 5-15\% of 2MASS detections in the magnitude range of interest for our catalog (7 $<$ H $<$ 13) are located within 6\arcsec of another 2MASS source, and may thus may not be resolvable into separate sources in WISE imaging.  Neighboring sources are typically 2-4 magnitudes fainter in $H$ than the candidate target, however, suggesting that the merged WISE counterpart will likely be dominated by the emission from the $7 < H < 13$ 2MASS source, providing a reasonably accurate description of the source's spectral energy distribution and avoiding a false/spurious identification as a potential IR excess source.  Figure \ref{fig:noCounterpartsSpatial} compares the spatial distributions of Spitzer identified YSOs in Orion that do and do not possess WISE counterparts, showing that sources lacking WISE counterparts are preferentially found in sub-regions of Orion with high stellar number densities, as well as high extinctions and elevated IR backgrounds from associated nebulosity. These factors, and visual inspection of a representative sample of the sources that lack WISE counterparts, point to the combined influence of crowding and elevated backgrounds in suppressing the recovery of WISE counterparts for the YSOs cataloged by \citet{Megeath2012}, and by extension, the remainder of our catalog. 

Aside from a complete lack of a WISE catalog counterpart, the presence of low-quality WISE photometry is the second most common reason that bona fide YSOs in the \citet{Megeath2012} catalog fail to be recovered by the KL14 selection algorithm. As described in Sec. \ref{sec:KL14explain}, the KL14 selection method applies quality cuts to WISE photometry such as those shown in Figure \ref{fig:qualitycuts}, which accounts for 523 of the YSOs in the \citet{Megeath2012} catalog which are not recovered in our parsing of the WISE catalog.  A final 697 YSOs from the \citet{Megeath2012} catalog are excluded from a YSO classification by the KL14 algorithm based on conservative cuts in W1 vs W1 - W2 color-magnitude and (W1 - W2) vs (W3 - W4) color-color space to eliminate extragalactic SFGs and AGN, or as potential Galactic AGB stars. This count also includes sources lost due to failing the tests for Transition Disk classification as defined in KL14. As this test shows, these cuts are conservative, and exclude some bona fide YSOs. We retain the cuts in generating our sample of WISE-identified candidate YSOs, however, choosing to sacrifice some likely YSOs to maintain a lower level of contamination across our full sample. Both Figure \ref{fig:megeathFlowchart} and Table \ref{tab:megeathResults} summarize the results of applying the multidimensional selection to the \citet{Megeath2012} catalog and demonstrate the number of YSOs recovered with our selection.

%% ----- Megeath results table -----
%results for Megeath as run through the Koenig selection. 
\begin{deluxetable*}{lcccc}
\tablewidth{\linewidth}
\tabletypesize{\scriptsize}
\tablecaption{Megeath YSOs Successfully Selected by KL14 Algorithm\label{tab:megeathResults}}
\tablehead{ 
\colhead{} & \colhead{Total Sources} & \colhead{Sources}  & \colhead{Sources} & \colhead{Sources} \\
\colhead{Test} & \colhead{Tested} & \colhead{Still Viable}  & \colhead{Eliminated} & \colhead{Classified} 
}
\startdata
WISE counterpart&3478&2470&1008&--\\
Quality Cut&2470&1947&523&--\\
SFG Test&1947&1870&77&--\\
AGN Test&1870&1465&405*&--\\
\hline
WISE Class I&1465&1315&--&150\\
WISE Class II&1315&302&--&1013\\
2MASS Test&302&215&--&--\\
2MASS Class I&87&67&--&20\\
2MASS Class II&67&--&--&67\\
Transition Disk&215&--&201&14\\
\hline
w4 Class 1&405&--&400&5\\
AGB Test&1269&1250&19&--\\
\hline
Total Candidates &3478&--&2228&1250\\
%\hline 
%Total & number  of Pleiades Stars Detected & 1334 & (82.3\%) \\
%Total & number  w/ $<$50 visits: & 287 & (17.7\%)
\enddata
\tablecomments{*These sources are not eliminated until they are passed through the W4 Class I test. }
\end{deluxetable*}

\subsection{Pan-STARRS-Variability Selected Young Stars} \label{sec:PanSTARRS}

To extend the uniform sample to include diskless Class III pre-main sequence stars, we utilize multi-epoch Pan-STARRS1 3$\pi$ \citep[PS1 3$\pi$; for more on the 3$\pi$ survey design, see sections 3.2 and 6 in][]{Chambers2016} photometry to identify optically variable pre-main sequence stars.  

\begin{figure}[t!] 
\centerline{\includegraphics[angle=0,width=\columnwidth]{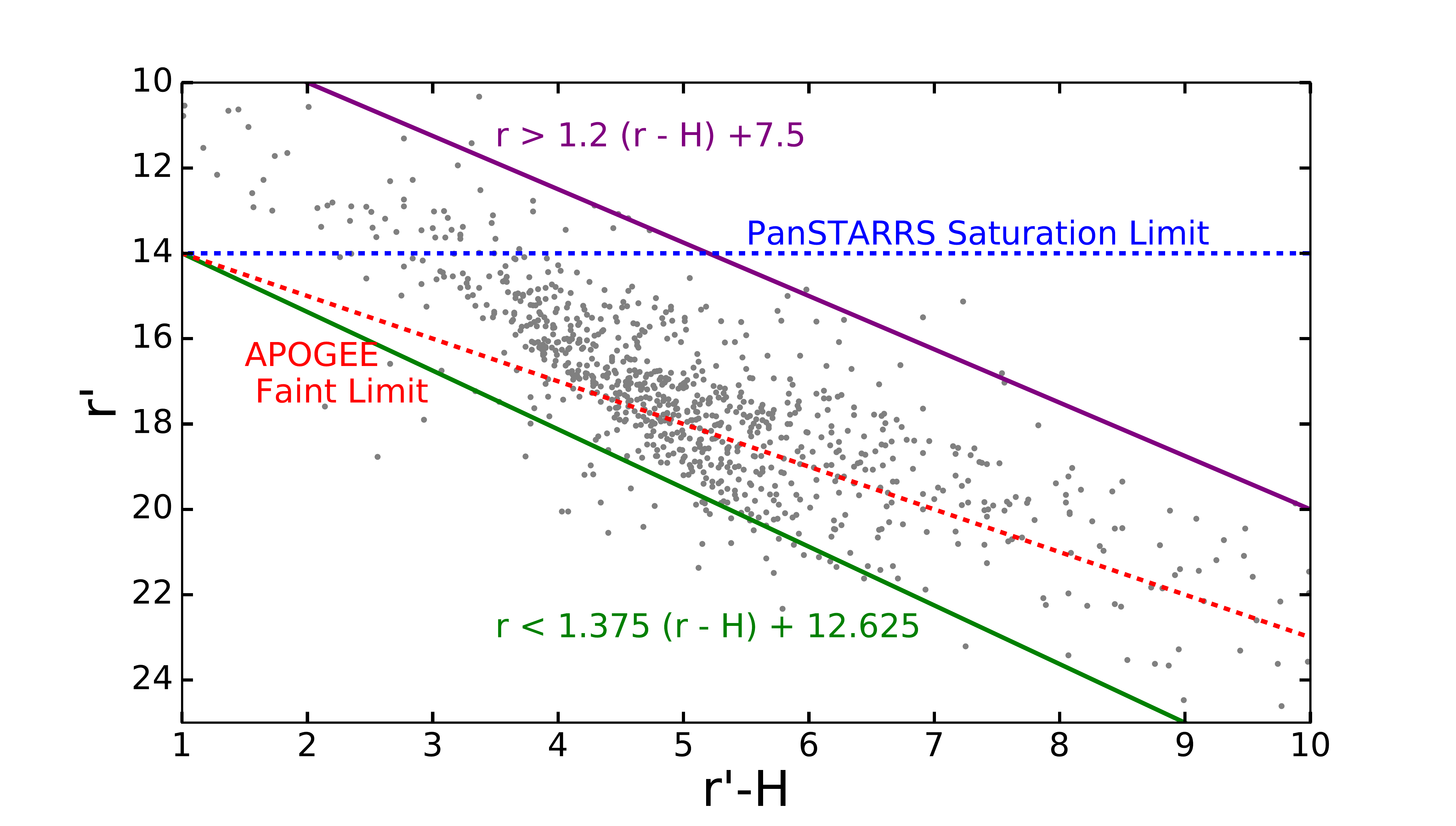}}
\caption{$r$-$H$ vs. $r$ color magnitude diagram of members identified by \citet{Fang2013}, along with color-magnitude cuts shown in green and purple that bracket these known members, which can be applied to Pan-STARRS1 photometry (not shown) to identify and select YSO candidates.   In blue and red are the limits for Pan-STARRS saturation and APOGEE faintness, respectively. \citet{Fang2013} sources mainly reside between the purple and green cuts. The query of the PS1 3$\pi$ dataset identified sources (omitted here for clarity) whose mean colors and magnitudes placed them within the green and purple lines.}
\label{fig:PanStarrsCutsFang}
\end{figure}

To select optically variable pre-main sequence members of the OSFC from the PS1 3$\pi$ catalog, we first imposed color-magnitude cuts informed by optical photometry of known YSOs in Orion compiled by \citet{Megeath2012} and \citet{Fang2013}. 
Based on the locations of YSOs in those catalogs, we extracted PS1 3$\pi$ sources that satisfied the following cuts:

$$1.71(g-H) + 4.85 < g < 1.85(g-H) + 10.14$$
$$1.2(r-H) + 7.5 < r < 1.375(r-H) + 12.625$$
 
\noindent These cuts are shown in Fig. \ref{fig:PanStarrsCutsFang}, along with the known YSOs cataloged by \citet{Fang2013}, which the cuts were designed to retain; the resultant sample of selected PS1 3$\pi$ sources are not shown, as they densely fill the allowed area. 
 
\begin{figure}[t!] 
\centerline{\includegraphics[angle=0,width=\columnwidth]{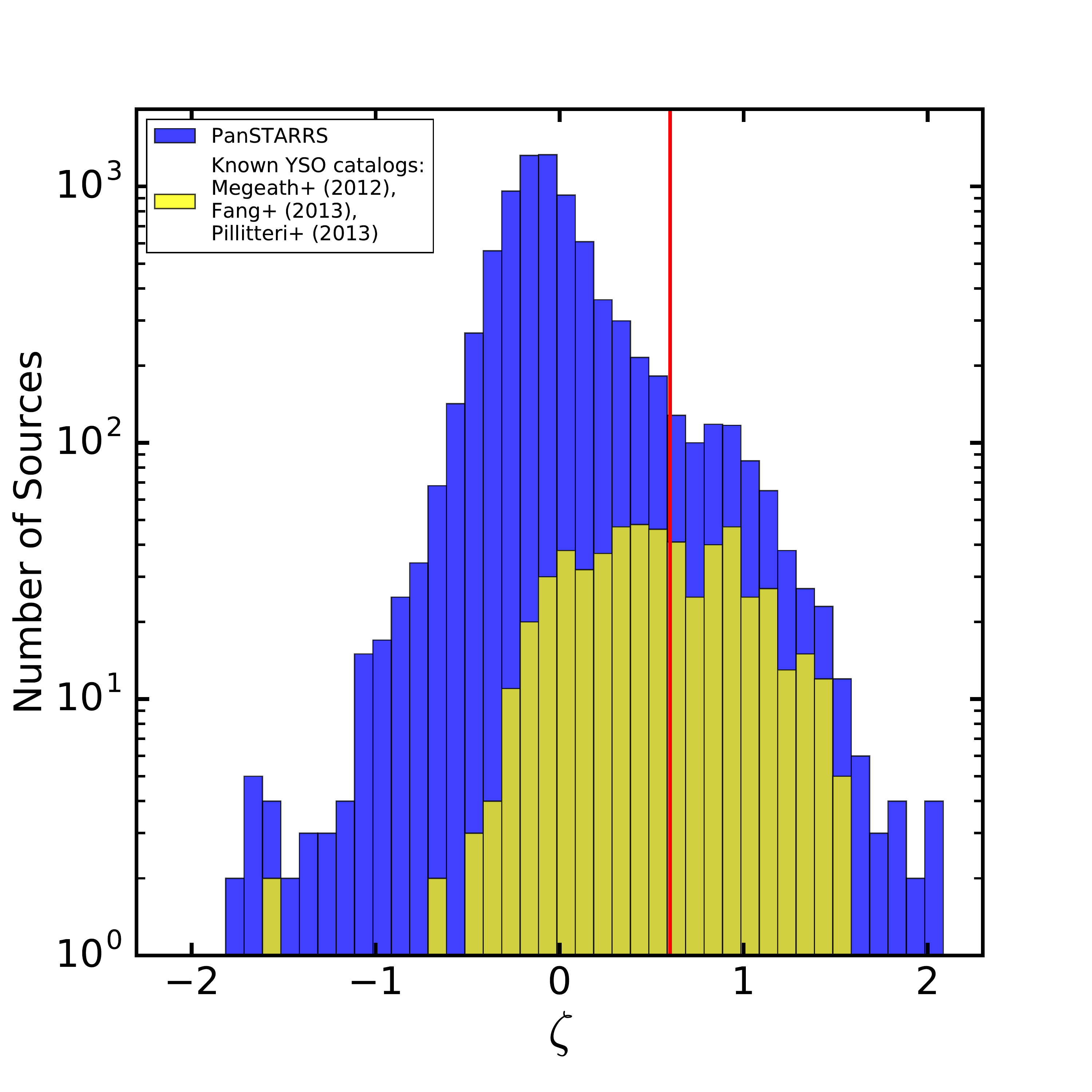}}
\caption{Histograms of $\zeta = \log ( \frac{\sigma_F}{err_F\sqrt{N}} )$ for all PS1 3$\pi$ sources in the OSFC footprint that meet the color-magnitude criteria in Fig. \ref{fig:PanStarrsCutsFang} (blue), and the subset of catalogs previously identified as a YSOs as identified by \citet{Fang2013}, \citet{Megeath2012}, and \citet{Pillitteri2013} (yellow). The red line represents the 3-sigma variability cut; for values of $\zeta$ larger than this limit, YSOs represent a substantial fraction of the sample.}
\label{fig:VariabilityHist_I}
\end{figure}
 
The PS1 3$\pi$ survey provides multi-band, multi-epoch $grizY$ photometry for millions of stars, QSOs and galaxies, but the time domain is not densely sampled: in the PS1 data analyzed here, a typical source  was observed $\sim$8 times in each filter, with a typical timescale between observations of 100 days to a year, but with some observations spaced more closely in time \citep[for more on the PS1 cadence, see section 3.2.4 by][]{Chambers2016}.  Sparse light curves with non-simultaneous color information  are not well-suited to metrics designed to detected correlated multi-band variability, such as the Welch-Stetson statistic \citep{Welch1993,Stetson1996}.  We instead identified bona-fide variables using a relatively simple, logarithmic variability metric computed by comparing a source's observed variability to its median photometric precision as reported by the PS1 pipeline: 
$$\zeta = \log \left( \frac{\sigma_F}{err_F\sqrt{N}} \right)$$ 
\noindent where $\sigma_F$ is the observed standard deviation in measurements over all epochs, $err_F$ is the expected error per epoch and $N$ is the number of epochs in all PanSTARRS filters\footnote{This metric was originally intended to utilize a filter-specific value of N, but due to a coding error, the total number of observations across all PanSTARRS filters was used in the calculation on which target selection was based. This implementation nonetheless preserves $\zeta$'s utility as a relative variability indicator, however, particularly as calibrated against values of $\zeta$ measured for known YSOs. We therefore report the results from the $\zeta$ values calculated in this way, and as used in the target selection process.}. Computing this metric for each of the PS1 bands ($grizY$), we compared the $\zeta$ values for the full sample of PS1 sources to those measured for previously identified Orion members; Figure \ref{fig:VariabilityHist_I} shows this comparison for $\zeta$ values for PS1 $i$-band detections. As Figure \ref{fig:VariabilityHist_I} indicates, no value of $\zeta$ can be used to select a sample composed only of known Orion members. Above a threshold of $\zeta = 0.6$, however, which represents a $\sim$3$\sigma$ detection of variability, known YSOs do represent a significant fraction of the total sample.  Conversely, below the $\zeta=0.6$ threshold known Orion members quickly contribute only a modest fraction of the sources at each value of $\zeta$. 

\begin{figure}[t!] 
\centerline{
\includegraphics[angle=0,width=\columnwidth]{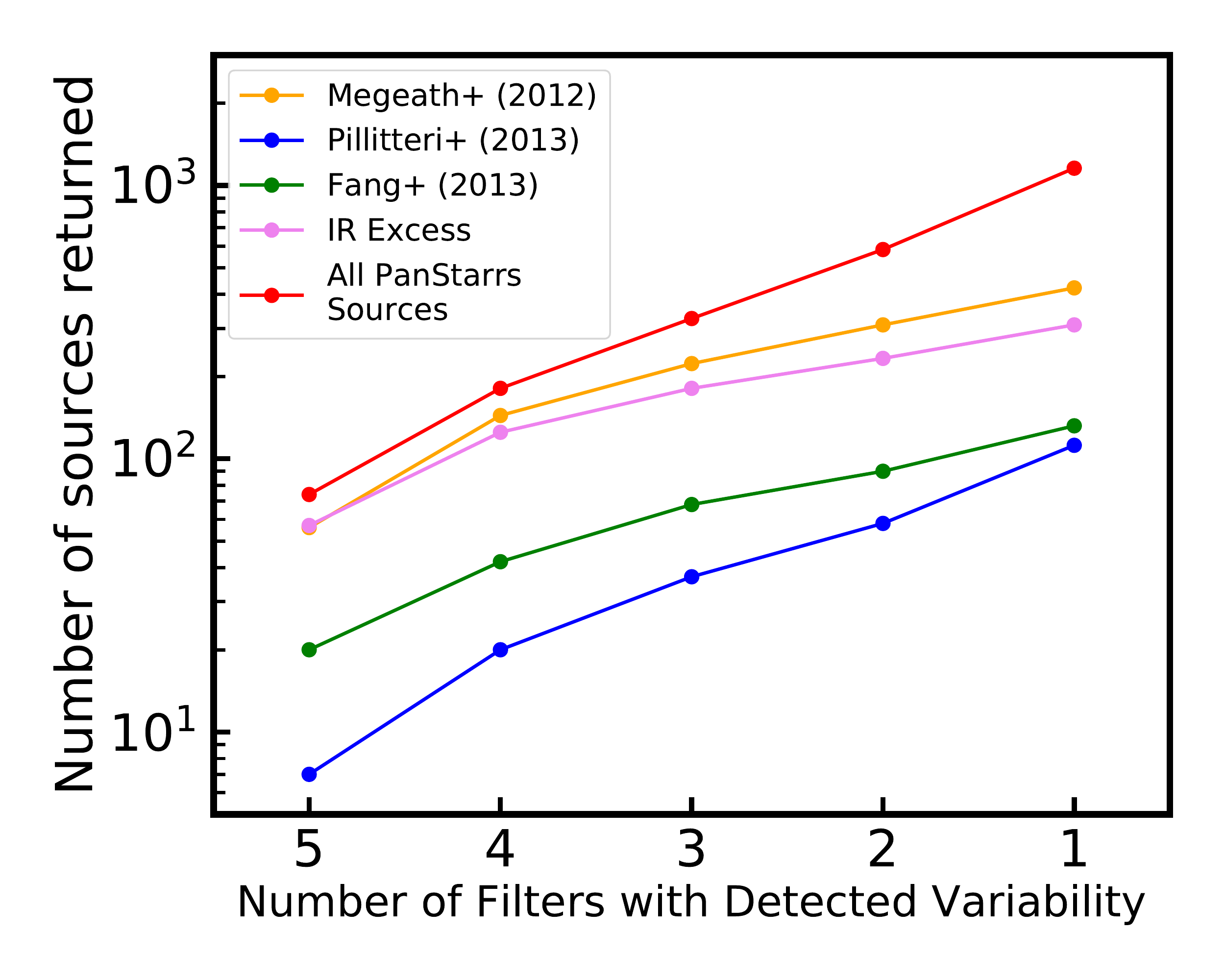}}
\caption{Numbers of sources in the Orion A region that are flagged as photometrically variable from PanSTARRS data as a function of the number of filters the variability must be detected in. Catalogs of known YSOs begin to level off near three filters, while the number of sources selected from the PS1 3$\pi$ survey continues to increase as the number of filters required for a variability classification decreases.}
\label{fig:PanStarrsPoints}
\end{figure}

The contamination of non-OSFC members in a variability selected sample can be further reduced, however, by requiring sources to meet the $\zeta=0.6$ threshold in multiple filters: this requirement will eliminate sources whose apparent variability in one or a few filters stems from spurious or low S/N measurements, while sources with \textit{bona fide} intrinsic variability will meet the cut in all, or nearly all, filters with reliable measurements.  To determine the number of filters in which variability must be detected in order to classify a source as a likely YSO, we examined how the identification of known YSOs depends on the number of filters considered. Figure \ref{fig:PanStarrsPoints} compares the number of sources that meet the $\zeta$ threshold in $n$ or more filters, when selecting from the full (color-magnitude restricted) PS1 3$\pi$ sample or from subsets of previously cataloged Orion members.  The number of candidate variables extracted from the full PS1 3$\pi$ catalog drops by approximately an order of magnitude when the number of filters a source must exhibit variability in is raised from one to three; by contrast, the number of previously known members that are flagged as candidate variables only drops by a factor of 2 when raising the threshold from one filter to three. Further increasing the number of filters in which a source must exhibit variability to be considered a candidate YSO produces similar reductions in the samples of variables extracted from the full PS1 3$\pi$ catalog versus the subsets of known YSO members, and thus no gain in the ratio of \textit{bona fide} YSOs to non-YSO contaminants. We therefore conservatively adopted a 3-filter requirement for a source to be classified as a bona fide variable and a candidate pre-main sequence star.  We also investigated potential biases due to color dependences in the photometric variability of known YSOs, and see no clear color dependencies in our sample.  Aside from a modest ($\sim$15\%) enhancement in the fraction of YSOs which are identified as variables in the $r$ filter, potentially due to variations in H$\alpha$ line emission from either chromospheric activity or mass accretion, all other filters flag a consistent 52-57\% of YSOs as photometric variables. 

Requiring variability detections in three filters is an attempt to balance the competing demands of identifying as many candidate members as possible, while also reducing contamination of the sample by photometrically variable non-members.  Based on the tests shown in Figure \ref{fig:PanStarrsPoints}, we expect that nearly 50\% of the pre-main sequence stars in this magnitude range are properly classified as variable by our selection technique. We cannot eliminate contamination completely, however, and as such we expect that our final sample of optically variable candidate pre-main sequence stars still contains a substantial numbers of other types of photometric variables, such as background AGB stars. 

\begin{figure*}[t!] 
\centerline{\includegraphics[angle=0,width=5in]{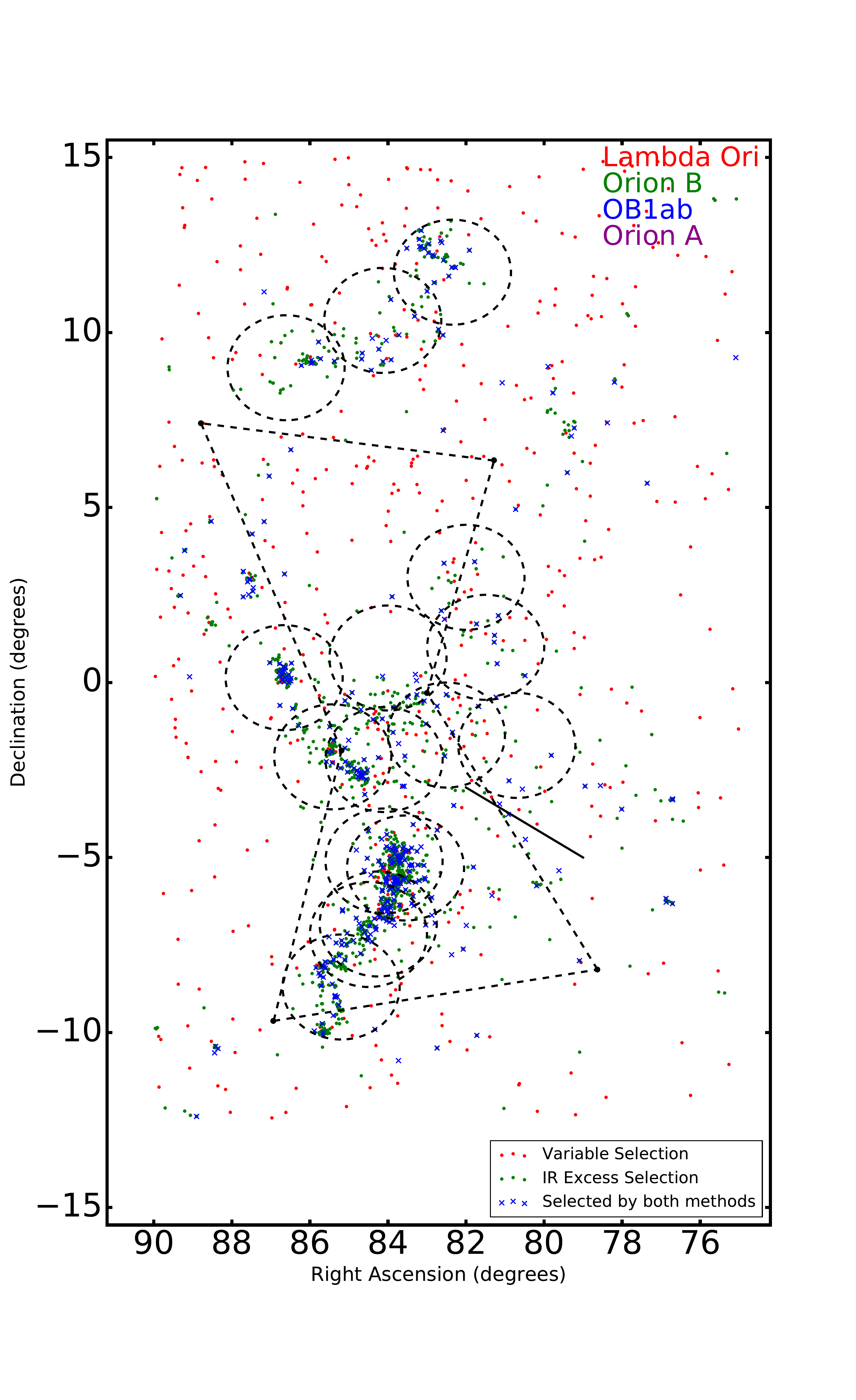}}
\caption{Spatial distribution of candidate YSOs selected via 2MASS+WISE photometry and PS1 3$\pi$ multi-epoch photometry, relative to the plate boundaries of the APOGEE-2 YSO Orion program.}
\label{fig:PanSTARRSOrion}
\end{figure*}

\section{Selected Targets and Final Plate Designs in Orion}\label{sec:supplementalSources}

The presence or absence of strong selection effects in the final sample observed in this survey will strongly influence the types of questions that the survey can address, and the complexity of the analysis that must be conducted to correct for those selection effects.  As a result, in assembling our final target lists and plate designs, we assigned the highest priority to objects in the `uniform sample' constructed from the union of the IR \& variability selected catalogs whose construction was described in Section \ref{sec:uniform}.  In Sec. \ref{sec:AllOfOrion} we describe the process by which these IR \& variability selected sources were included in the APOGEE-2 Orion plate designs; in Sec. \ref{sec:finalPlates}, we describe how additional sources were added to these designs to fill the remaining optical fibers, providing a larger sample of bona fide members, albeit with a more complex and heterogeneous selection function. 

\subsection{A Uniform Sample of IR/Variability Selected YSOs Spanning All APOGEE-2 Orion Fields}\label{sec:AllOfOrion}

Applying the KL14 algorithm as described in Section \ref{sec:KL14explain} to sources in the AllWISE catalog within a $\sim$400 square degree region bounded by $75 <$ R.A. $<90~\deg$ and $-12.5 <$  Dec. $< 15~\deg$, we identified an initial sample of 2699 candidate YSOs throughout Orion. The resulting catalog of YSOs with evidence of circumstellar disks is given as Table \ref{tab:IRmembership} of the appendix and shown in green in Figure \ref{fig:PanSTARRSOrion}, with the locations of each APOGEE field indicated for reference. For targeting purposes, we restrict this catalog to sources with $H<$12.4, where the combined spectrum for a six visit source delivers S/N $\sim$100 and enables the measurement of robust stellar parameters such as T$_{eff}$ and log $g$ \citep[see Fig. 22 by][for APOGEE's S/N performance as a function of magnitude for 6 visit sources]{Nidever2015}. Restricting the catalog to sources with $H<$12.4 produces a sample of 1307 likely IR excess YSOs throughout the Orion complex. Over-densities associated with known sub-populations of Orion are visible in Figure \ref{fig:PanSTARRSOrion}; the most prominent condensations are those associated with the ONC, Orion A \& B clouds, and the $\sigma$ Ori cluster; more diffuse populations such as the $\lambda$ Ori and Orion OB1a/b associations are also visible, albeit at somewhat lower contrast. 

Selecting sources in the color-magnitude filtered PS1 3$\pi$ catalog that exhibited photometric variability (i.e. $\zeta \geq 0.6$) in three or more filters identifies 3697 likely variables within the same $\sim$400 square degree footprint used for the identification of IR excess sources. All 3697 likely variables are listed in Table \ref{tab:PSmembership2} of the appendix, but as with the IR selected candidates we target only those with $H<$12.4.  This identifies 990 potential APOGEE-2 targets across the full footprint of the OSFC, which are shown in red in Figure \ref{fig:PanSTARRSOrion}. 

\subsection{Merging Targets to Produce Final Plate Designs} \label{sec:finalPlates}

In most APOGEE-2 Orion Survey fields, the catalog of $H \leq$12.4 uniformly selected candidate YSOs does not include all previously identified members of the OSFC, nor does it fill all of the available science fibers on a given plate design, particularly after resolving conflicts between multiple sources with separations less the APOGEE-2N spectrograph's 72\arcsec fiber collision radius. As a result, after assigning fibers to $H \leq$12.4 uniformly selected candidate YSOs on each of the plates covering a given field, we assigned additional targets from a prioritized list of previously confirmed members and other candidate YSOs. Due to the restricted footprints of existing membership catalogs, we provide a brief overview of the overall prioritization scheme, followed by a detailed discussion of the membership catalogs and selection processes most relevant to each of Orion's sub-regions. 

\subsubsection{Prioritization Scheme}

We make use of a bitmask consisting of binary flags to store and sort the relative priorities assigned to each of the potential targets in this program. Targets within a given membership catalog, or satisfying the criteria associated with a given selection algorithm, are assigned a bit corresponding to a binary value of 2$^n$; the value of the priority flag used for each catalog is outlined in Table \ref{tab:OrionPriority}. We then determine the prioritization of all targets within a given field according to the sum total of all the binary priority flags that have been set for each source. In this system, sources present in catalogs corresponding to higher values of $n$ will be assigned to fibers first; in this way, the value of the bit assigned to a catalog serves mainly to order the input catalogs according to our qualitative assessment of their potential contribution based on several factors, including the total size of the catalog, as well as the completeness and contamination of its described sample. Among sources with the same bit set, any bits set for lower priority catalogs will provide a slight increase to the value of the total priority value, and cause that source to be targeted before sources that are absent from the lower priority catalogs. 

The prioritization method we have adopted ensures that our uniformly selected sources do not compete for fibers with other, less homogeneously selected members, and thus protects the simplicity of the uniform sample's selection function.  As an example, a source in the Orion A cloud that was included in the \citet{Megeath2012} and \citet{Pillitteri2013} catalogs, and also selected as a likely YSO by our implementation of the KL14 method, would be assigned bitwise priority flags of 2$^5$, 2$^6$, and 2$^8$, and would have a final priority value of 2$^8$ + 2$^6$ + 2$^5 =$ 352. A source that was only selected by our KL14 implementation would have a final priority value of 2$^8$, whereas a source that was not selected in our KL14 implementation but was included in the catalogs compiled by \citet{Megeath2012} and \citet{Pillitteri2013} would have a final priority of 2$^6$ + 2$^5 =$ 96.  Applying our prioritization scheme to these three sources, the source detected in all three catalogs would be targeted first, followed by the KL14-only source, and finally by the source detected by \citet{Megeath2012} and \citet{Pillitteri2013}. To ensure our most consistent selection methods are applied across Orion's full footprint, we assign the highest priority values to the candidate YSOs we identify on the basis of their 2MASS+WISE and PS1 3$\pi$ photometry.  Due to a miscommunication among the team, different maximum bit values were used for the Orion OB1a/b and $\lambda$-Ori/Orion AB regions: 2$^8$ for the $\lambda$-Ori and Orion AB regions, and 2$^{10}$ for the Orion OB1a/b regions. Nonetheless, the 2MASS+WISE and PS1 selected targets were assigned the maximum and penultimate bit values in each region's ranking scale, so the relative prioritization of all target classes remains consistent across the full OSFR. Table \ref{tab:OrionPriority} documents the various catalogs and selection criteria from which we draw targets in our Orion survey.

\subsubsection{$\lambda$ Ori}

\begin{figure}[h] 
\centerline{\includegraphics[angle=0,width=\columnwidth]{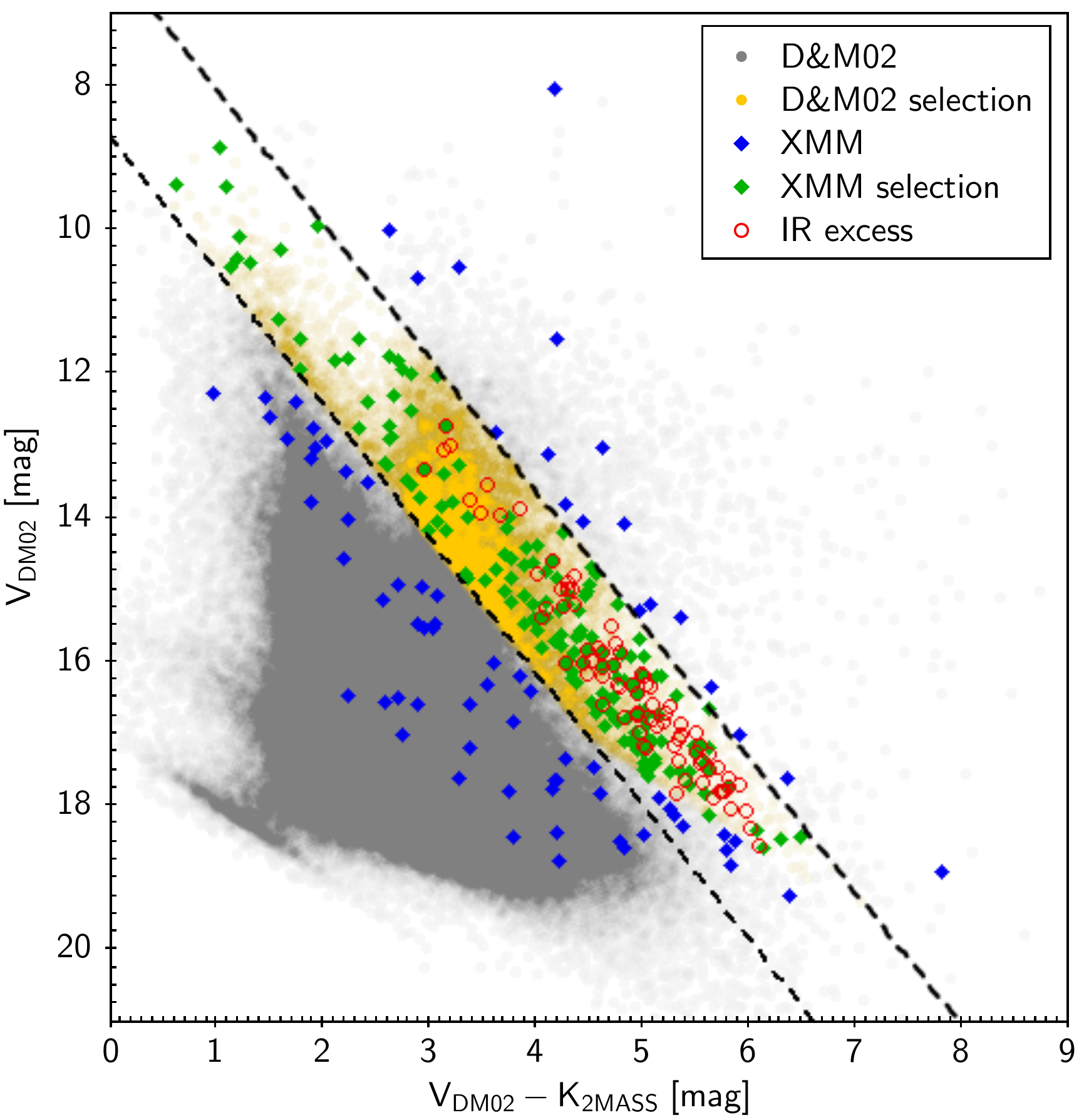}}
\caption{$V$ vs. $V-K$ color-magnitude diagram of sources in the $\lambda$ Ori field from the catalog of Dolan and Mathieu \citet{Dolan2002}. Known members of the association (XMM X-ray and infrared excess sources) trace the empirical locus defined by the two dashed lines, from which we selected additional sources to fill vacant fibers in APOGEE plate designs. }
\label{fig:LambdaOriPhotSelect}
\end{figure}

As one of the older and sparser regions within Orion, $\lambda$ Ori yielded only 147 H$<$12.4 mag candidate YSOs selected by our 2MASS+WISE and PS1 3$\pi$ selection methods across three distinct APOGEE-2 fields.  To fill the remaining fibers in these fields, we first targeted nearly 400 confirmed cluster members identified by \citet{Dolan2001} and \citet{Hernandez2010} on the basis of optical spectroscopy and mid-infrared excesses, respectively. The catalogs produced by \citet{Barrado2007} from Spitzer IRAC photometry and \citet{Barrado2011} yielded no new H $<$ 12.4 candidates; eight additional sources were identified and targeted from the XMM-Newton-based catalog compiled by \citet{Franciosini2011}. The remaining free fibers were filled with 1257 (primarily single-visit) candidate $\lambda$ Ori members selected as follows: we defined an empirical locus in the optical-infrared $V$ vs. $V-K$ color-magnitude diagram traced by members from the \citep{Hernandez2010} catalog as well as high likelihood candidates in the 3XMM-DR4 source catalog \citep{Rosen2015} and sources with WISE+2MASS colors indicative of a mid-IR excess ($(K-W2)>$0.5), as illustrated in Figure \ref{fig:LambdaOriPhotSelect}. From this locus we drew 43 X-ray sources from the \citeauthor{Rosen2015} catalog; five candidate members identified by \citet{Koenig2015} via a preliminary version of the KL14 infrared selection method; two \citet{Dolan2002} sources mid-IR excess and 1207 (relatively low likelihood) candidates selected from the \citet{Dolan2002} photometric catalog as sources lying within the color-magnitude diagram locus highlighted in Figure \ref{fig:LambdaOriPhotSelect}.

\subsubsection{Orion OB1a/b}
The Orion OB1 association is one of the largest and nearest sites of recent star formation, and includes the $\sim7 - 10$ Myr old OB1a and $\sim4-6$ Myr old OB1b populations \citep{Briceno2005}. The targets selected for the six Orion OB1ab plates include all H$< 12.4$ YSOs from our 2MASS+WISE and PS1 3$\pi$ selection (178 and 115, respectively). We also included 153 X-ray sources from the 3XMM-DR5 Source catalog \citet{Rosen2016} with 2MASS counterparts that lie within the OB1ab fields. Most of these X-ray sources are located within the Ori OB1a/b-A, Ori OB1a/b-E and Ori OB1a/b-F plates (see Fig. \ref{fig:sky_plates}). To sample the brighter end of the cluster sequence, we selected 108 highly probable stellar members of Orion OB1 from the \citet{Kharchenko2005} catalog. These sources populate all plates except Ori OB1a/b-A (see Fig. \ref{fig:sky_plates}). To fill the plates with bona fide members, we considered several studies to include 381 spectroscopically confirmed stellar members, which are mainly focused in the $\sigma$ Ori cluster and the 25 Orionis stellar group: 130 and 56 members from \citet{Briceno2005,Briceno2007} respectively, 178 from \citet{Hernandez2014}, 11 from \citet{Downes2014} 

%%% ----- Priority Flag table -----
%describes the construction of priority flags for each source.
\startlongtable
\begin{deluxetable*}{lcccc}
\tablewidth{\linewidth}
\tabletypesize{\footnotesize}
\tablecaption{Priority Flag\label{tab:OrionPriority}}
\tablehead{ 
\colhead{Selection} & \colhead{Priority} & \colhead{Sources} & \colhead{Sources}  & \colhead{Primary} \\
\colhead{Source} & \colhead{Bit} & \colhead{Flagged\tablenotemark{a}} & \colhead{Assigned\tablenotemark{b}}  & \colhead{Assignment\tablenotemark{c}}  
}
\startdata
\multicolumn{5}{c}{\emph{Full Footprint\tablenotemark{d}}} \\
WISE+2MASS IR excess (KL14 algorithm) & 2$^8$ & 1307 & 1034 & 1034 \\
Pan-STARRS variability & 2$^7$ & 990 & 531 & 212 \\
\hline
\multicolumn{5}{c}{\emph{$\lambda$ Ori}} \\
WISE+2MASS IR excess (KL14 algorithm) & 2$^8$ & 143 & 123 & 123 \\
Pan-STARRS variability & 2$^7$ & 73 & 62 & 24 \\
\citet{Hernandez2010} & 2$^6$ & 291 & 285 & 276 \\
\citet{Dolan2001} & 2$^5$ & 177 & 164 & 117 \\
\citet{Barrado2007} & 2$^4$ & 48 & 46 & 0 \\
\citet{Franciosini2011} & 2$^3$ & 50 & 48  & 8 \\
\citet{Barrado2011} & 2$^3$ & 45 & 45 & 3 \\
\citet{Rosen2015} XMM & 2$^2$ & 191 & 177 & 43 \\
\citet{Koenig2015} & 2$^2$ & 156 & 136 & 5 \\
\citet{Dolan2002} + WISE & 2$^2$ & 61 & 53 & 2\\
\citet{Dolan2002} + CMD & 2$^1$ & 2984 & 1572 & 1206 \\
\hline
\multicolumn{5}{c}{\emph{Orion OB1a/b}} \\
WISE+2MASS IR excess (KL14 algorithm) & 2$^{10}$ & 240 & 177 & 177 \\
                Pan-STARRS variability & 2$^9$ & 165 & 137 & 57 \\
\citet{Megeath2012} Spitzer IR excess & 2$^8$ & 105 &  35 &  5 \\
  			  \citet{Kharchenko2005}  & 2$^8$ & 109 & 109 & 108 \\
   			  \citet{Briceno2005}     & 2$^8$ & 144 & 137 & 130 \\
   			  \citet{Briceno2007}     & 2$^8$ &  56 &  56 &  56 \\
			  \citet{Hernandez2014}   & 2$^8$ & 308 & 200 & 178 \\
              \citet{Hernandez2007}   & 2$^8$ &  91 &   2 &   2 \\
   			  \citet{Downes2014}      & 2$^8$ &  15 &  11 &  11 \\
     		  \citet{Bouy2014}        & 2$^8$ &  67 &  41 &   3 \\
              \citet{Suarez2017} & 2$^8$ &  18 &   6 &   6 \\
  						         XMM  & 2$^8$ & 314 & 237 & 172 \\
       			USNO+2MASS selection  & 2$^8$ & 4149 & 4149 & 2269 \\
            \citet{Broos2013} Chandra & 2$^6$ & 101 & 100 & 62 \\
              \citet{Caballero2008}   & 2$^5$ & 338 & 179 & 161 \\ 
\hline
\multicolumn{5}{c}{\emph{Orion A\&B}} \\
WISE+2MASS IR excess (KL14 algorithm) & 2$^8$ & 857 & 738 & 738 \\
Pan-STARRS variability & 2$^7$ & 406 & 355 & 132 \\
\citet{Megeath2012} Spitzer IR excess& 2$^6$ & 1345 & 1109 & 445 \\
\citet{Pillitteri2013} X-ray YSO & 2$^5$ & 492 & 446 & 287 \\
\citet{Broos2013} Chandra & 2$^4$ & 209 & 144 & 43 \\
\citet{Caballero2008} & 2$^3$ & 194 & 77 & 59 \\
\citet{Hernandez2007} & 2$^2$ & 222 & 78 & 23 \\
\citet{Hernandez2014} & 2$^1$ & 400 & 231 & 139 \\
\citet{Briceno2005} & 2$^0$ & 16 & 11 & 11 \\
\citet{Bouy2014}        & 2$^0$ & 2 & 2 &  2 \\
    XMM  & 2$^0$ & 115 & 25 & 5 \\
USNO+2MASS Selection & 2$^0$ & 400 & 271 & 216 \\
\enddata
\tablenotetext{a}{`Sources flagged' indicates how many sources within each region were identified by a particular selection method/membership catalog.}
\tablenotetext{b}{`Sources assigned' indicates how many sources within each region were identified by a particular selection method/membership catalog and assigned fibers after resolving potential fiber collisions.}

\tablenotetext{c}{`Primary assignment' indicates the number of sources within each region that were identified by a particular selection method/membership catalog and were assigned a fiber, but were not identified by any higher priority catalogs, such that the given catalog is `responsible' for their targeting.}

\tablenotetext{d}{Full Footprint values do not correspond to the sum of the values reported for fields spanning individual sub-regions. Full Footprint `Sources Flagged' are drawn from a much larger effective area, while targets in the overlap region between fields assigned to different sub-regions [i.e., OrionB-B and OrionOB1ab-A] are included in the `Sources Assigned' and `Primary Assignment' fields for multiple sub-regions.}
\end{deluxetable*}

\noindent and 6 from \citet{Suarez2017}. Additionally, we included 62 sources from the MYStIX catalog \citep{Broos2013}, 161 members from the study of \citet{Caballero2008}, 2 members from the list of \citet{Hernandez2007}, and 3 highly probable young stellar photometric candidates from \citet{Bouy2014}. The remaining fibers were assigned to 2269 photometric candidates selected according to their position in the $I$ vs $I-J$ diagram for the Ori OB1a/b-E and Ori OB1a/b-F plates where there is minimum extinction, and in the $I$ vs $I-K$ diagram for the rest of plates where the extinction is slightly greater (see Figs. \ref{fig:CMD_plate_A} and \ref{fig:CMD_plate_E} for examples of this selection method in plates Ori OB1a/b-A and E). These candidates were drawn from a photometric locus in each color-space, defined as a 0.15 mag wide space to both sides of an empirical isochrone traced by confirmed members and highly probable young stellar candidates from \citet{Kharchenko2005} as shown in Figures \ref{fig:CMD_plate_A} and \ref{fig:CMD_plate_E}.

\begin{figure}[h] 
\centerline{\includegraphics[angle=0,width=\columnwidth]{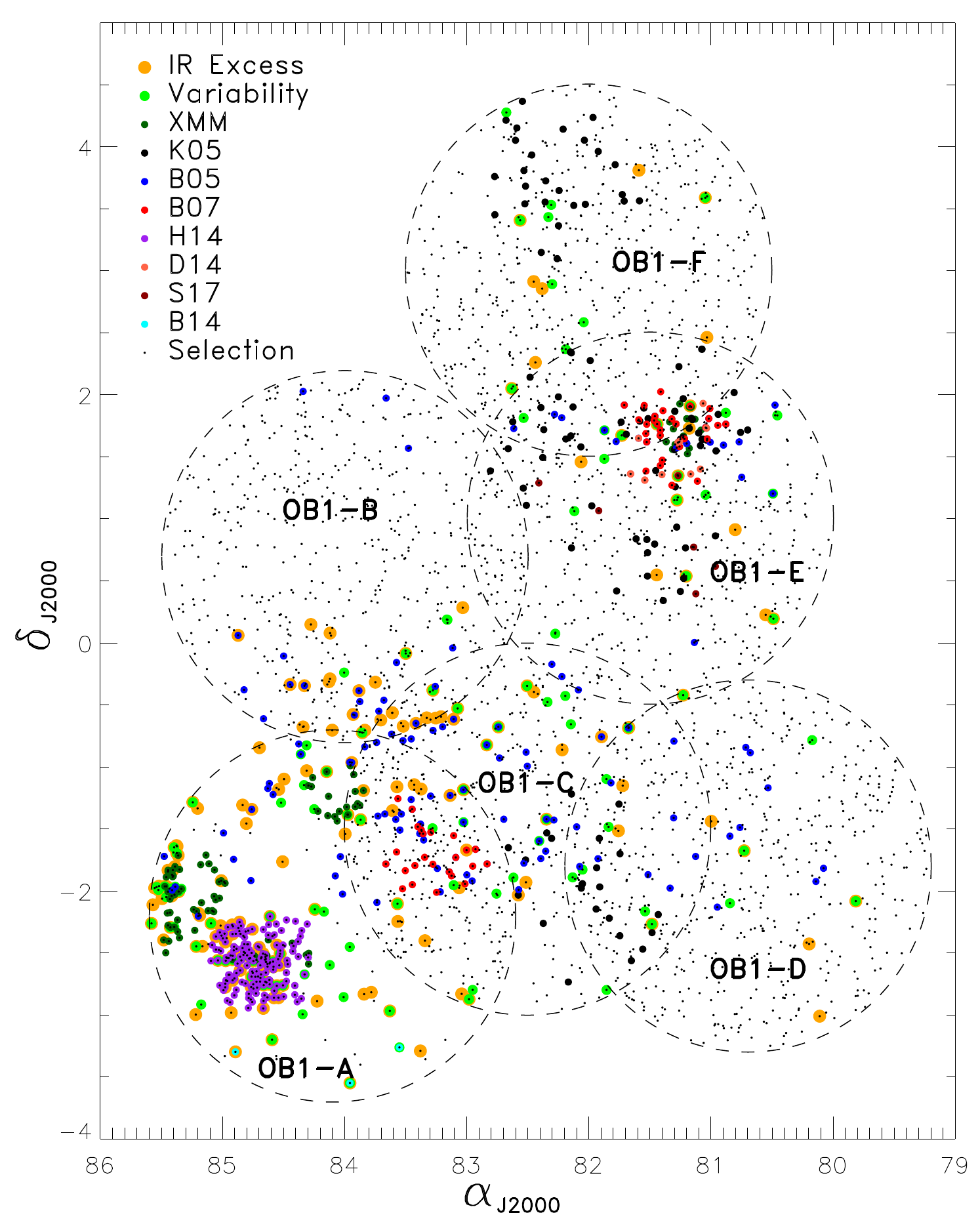}}
\caption{Spatial distribution of targets in the Ori OB1a/b plates. The color and size of the symbols indicate the source catalog and the priority of the targets, respectively. From top to bottom the abbreviations in the plot legend correspond to: 2MASS+WISE (IR excess), PS1 3$\pi$ (variability), 3XMM-DR4 catalog, \citet{Kharchenko2005}, \citet{Briceno2005,Briceno2007}, \citet{Hernandez2014}, \citet{Downes2014}, \citet{Suarez2017}, \citet{Bouy2014} and our USNO-2MASS selection. The gray points show the USNO+2MASS photometry inside the OB1a/b-A plate area.} 
\end{figure}
\label{fig:sky_plates}

\begin{figure}[h] 
\centerline{\includegraphics[angle=0,width=\columnwidth]{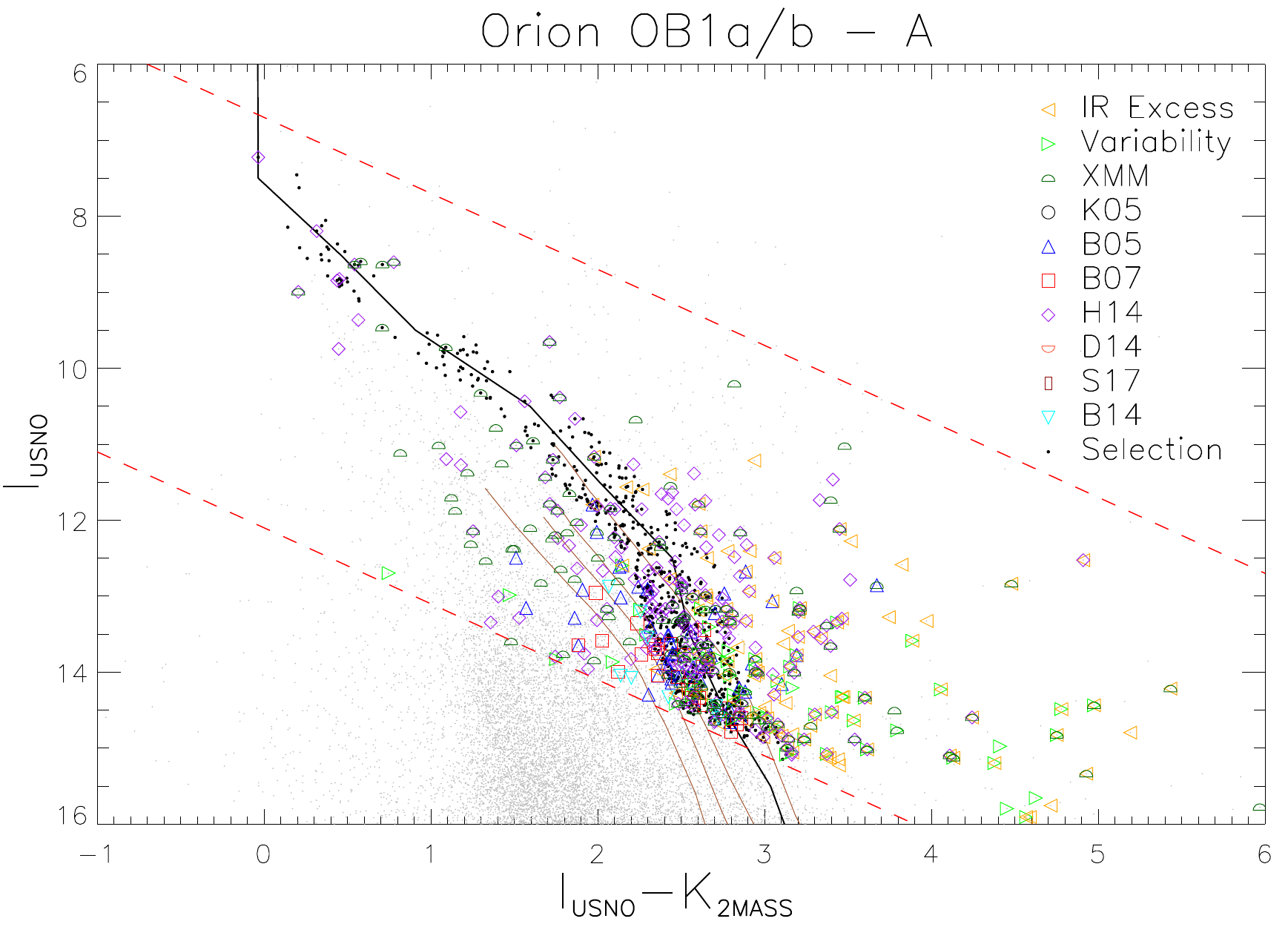}}
\caption{$I$ vs $I-K$ color-magnitude diagram of the selected targets for the APOGEE-2 Ori OB1a/b-A plate. The black solid curve corresponds to the empirical isochrone defined following the confirmed members and the high probability young stellar candidates, and the brown curves are the 1, 3, 5 and 10 Myr isochrones from \citet{Baraffe2015}. The red dashed lines indicate the APOGEE-2 limits. The labels and references are the same as in Figure \ref{fig:sky_plates}.}
\label{fig:CMD_plate_A}
\end{figure}

\begin{figure}[h] 
\centerline{\includegraphics[angle=0,width=\columnwidth]{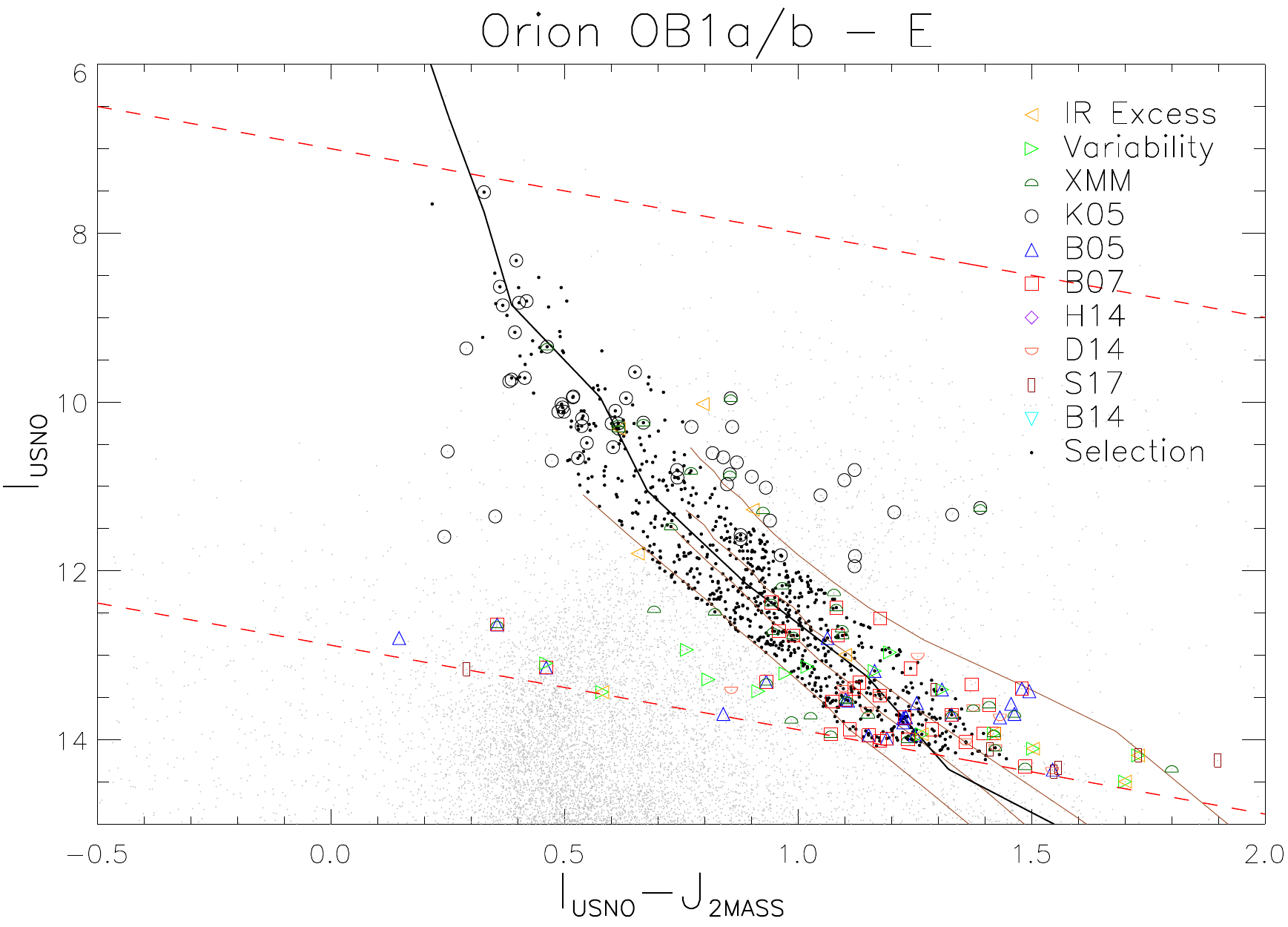}}
\caption{$I$ vs $I-J$ color-magnitude diagram of the selected targets for the APOGEE-2 Ori OB1a/b-E plate. All symbols and references are consistent with Figure \ref{fig:sky_plates} and Figure \ref{fig:CMD_plate_A}.}
\label{fig:CMD_plate_E}
\end{figure}

\subsection{Orion A \& B}

As the youngest and densest sub-regions within Orion, the Orion A \& B clouds have been extensively observed at IR and X-ray wavelengths, providing an extensive suite of existing membership catalogs and multi-wavelength observations to support target selection.  Nonetheless, nearly a thousand sources meet the criteria for our 2MASS+WISE and PanSTARRS selection in this region; combined with 450 and 288 sources drawn from the Spitzer and XMM based catalogs compiled by \citet{Megeath2012} and \citet{Pillitteri2013}, respectively, these catalogs provide more than 80\% of the targets we identify in these clouds. To these 1711 sources, we add 49 likely YSOs from $Chandra$ exposures of  of the Orion Nebula and Flame Nebula  \citep[][ Getman 2015, private communication]{Broos2013, Getman2014}, and 128, 62 and 144 confirmed or candidate $\sigma$ Ori members compiled by \citet{Caballero2008}, \citet{Hernandez2007} and \citet{Hernandez2014}, respectively.

%\begin{figure}[h] 
%\centerline{\includegraphics[angle=0,width=\columnwidth]{CollisionBias.eps}}
%\caption{Fraction of targeted sources within the Orion A \& B fields that were assigned a fiber on our final plate designs, as a function of the distance to their nearest neighbor (in arcseconds).  We achieve $\sim$90-100\% completeness for sources with nearest neighbor distances $>$72\arcsec, with sources lost only due to the overall limitation of 250 science fibers per plate.  Completeness drops for sources with smaller nearest neighbor distances, reaching a minimum of $\sim$60\% for sources with nearest neighbor distances $<$20\arcsec.  In regions this dense, however, there are typical 5 or more sources within a single fiber collision radius; we nonetheless obtain observations for more than half of these sources through the use of multiple plate designs for each field, allowing the survey to use repeated observations of the same field to build up a more complete sample in even the densest regions. }
%\label{fig:CollisionBias}
%\end{figure}

\section{Results: Evaluating YSO Yield with APOGEE Radial Velocities} \label{sec:RVvalid}

\subsection{Velocity distributions of cluster members and Galactic thin and thick disk stars in the $\lambda$ Ori A field}
To test the efficiency of our target selection, we compared the heliocentric radial velocities (RV) measured from targets in the early 2016 observations of the $\lambda$ Ori A field  to the RV distribution expected for the field population along the same line of sight. The RV distributions of the field populations from the Galactic thin and thick disks, as well as the Galactic halo, were simulated using the Besan\c{c}on Galaxy model \citep{Robin2003}. This model has been extensively tested within the sensitivity limits of 2MASS and appropriately reproduces the luminosity functions from the different components of the Galaxy \citep{Robin2003}. We assume the Besan\c{c}on model's standard Galactic parameters; that is, we compute RV distributions based on models that assume the structural and kinematic properties listed by \citet{Robin2003} in their Tables 1-4 for the young and old disk populations.

\begin{figure}[t!]
\centerline{\includegraphics[angle=0,width=\columnwidth]{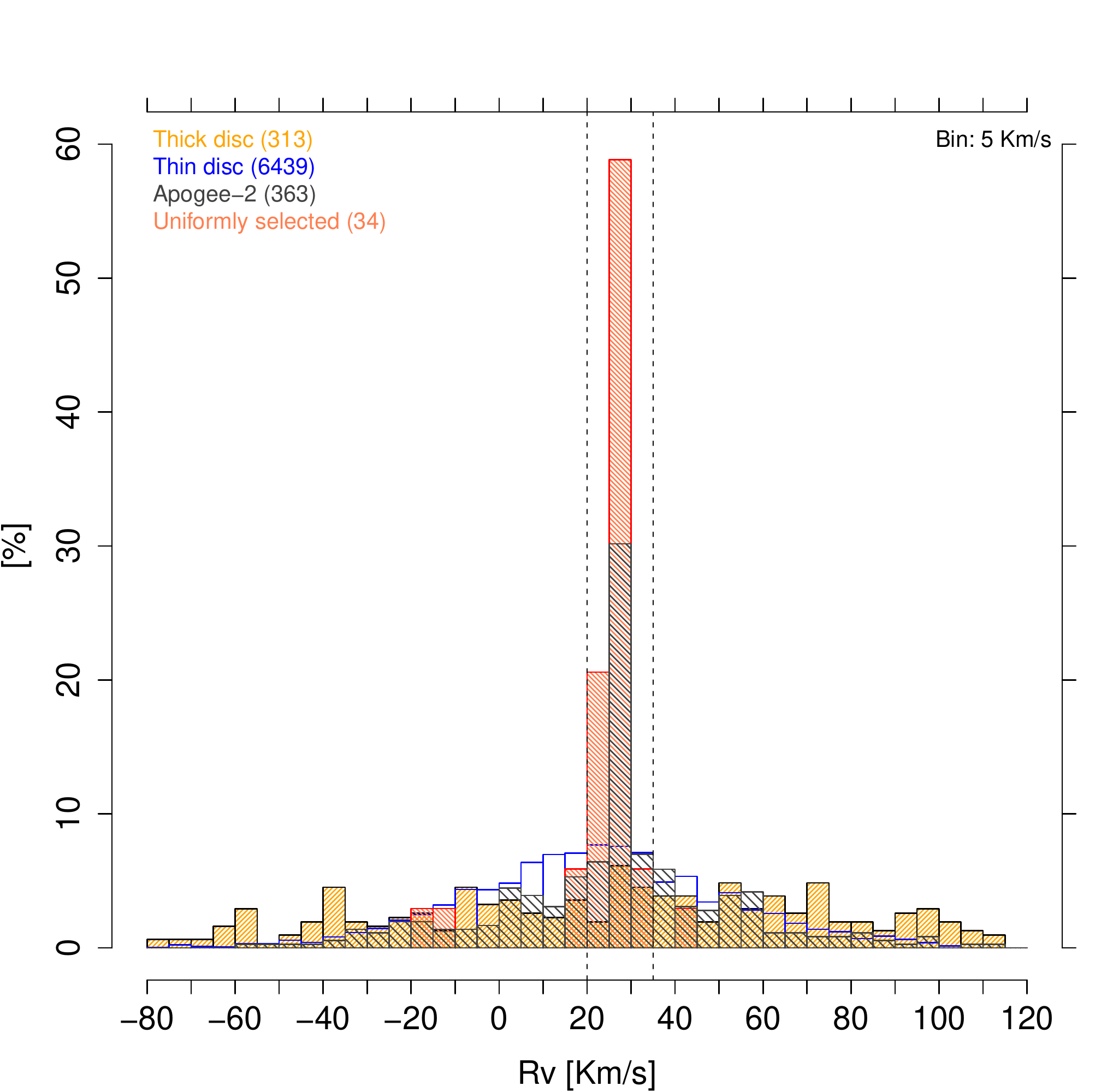}}
\caption{Distributions of heliocentric RVs along the line of sight to the $\lambda$ Ori-A field. The sample sizes in these distributions range from as small as 34 objects, up to nearly 6500: to place these on the same scale, the y-axis indicates the fraction of each distribution that falls within each bin. The black distribution shows all candidate YSOs observed with APOGEE-2 in the Lambda Ori A field; the red distribution shows the subset of those YSOs that were uniformly selected. Blue and orange distributions show respectively, the results from the simulations with the Besan\c{c}on Galaxy model for the thin and thick Galactic discs. Dashed lines indicate the range of the RVs measured for bona fide members of the major sub-populations of Orion: $\lambda$ Ori \citep{Dolan2001}, $\sigma$ Ori \citep{jeffries2006} and Orion A \citep{Tobin2009,DaRio2016}. The RV distributions of the thin and thick disk populations are significantly broader than those measured in Orion, such that $\leq$20\% of the Galactic contaminants accessible to our program would be expected to exhibit RVs within the 20-35 km/sec range we use to kinematically select candidate Orion members. Labels indicate the total number of stars in each sample and the bin size of the distributions. The distributions were normalized in order to emphasize their differences.}
\label{fig:vrad_besancon}
\end{figure}

The simulation was performed for the full area covered by the $\lambda$ Ori plate ($\sim$ 7 square degrees) and for a magnitude range ($7<H<12.8$) comparable to that of our selected targets. RV errors were assigned to each simulated source by drawing from a distribution matched to the RV precision of the APOGEE-2 observations by fitting an exponential function to the observed RVs of previously known members (which share a narrow range of intrinsic RVs) as a function of H-band magnitude. Figure \ref{fig:vrad_besancon} shows the RV distribution measured for APOGEE-2 targets in the $\lambda$ Ori-A field (plates 8879 and 8880), together with the simulated RV distributions for Galactic thin and thick disk stars along the same line-of-sight. The simulation predicts only two halo stars, which are not shown.

The RV distribution of the 363 targets observed with APOGEE-2 in the $\lambda$ Ori-A field during  early 2016 shows a strong peak, indicating the presence of a distinct kinematic population whose central velocity agrees with that previously measured for $\lambda$ Ori members \citep[24.5 km/sec; ][]{Dolan2001}. The Besan\c{c}on model predicts that the RV distributions of thin and thick disk stars have central velocities near that of the $\lambda$ Ori population, but much larger velocity dispersions, such that $\leq$20\% of the Galactic contaminants accessible to our program should exhibit RVs in the 20-35 km/s range occupied by previously confirmed Orion members.  Nonetheless, a substantial number of targets observed by APOGEE exhibit RVs more consistent with the field population than with cluster membership; this is not unexpected, particularly in the $\lambda$ Ori field, where many low-yield photometric candidates were added to complete plates after fibers had been assigned to the smaller samples of uniformly selected and previously confirmed members (see Table \ref{tab:OrionPriority}).

\subsection{Velocity distribution of uniformly selected targets in all observed fields}

\begin{figure*}[t!] 
\centerline{\includegraphics[angle=0,width=5in]{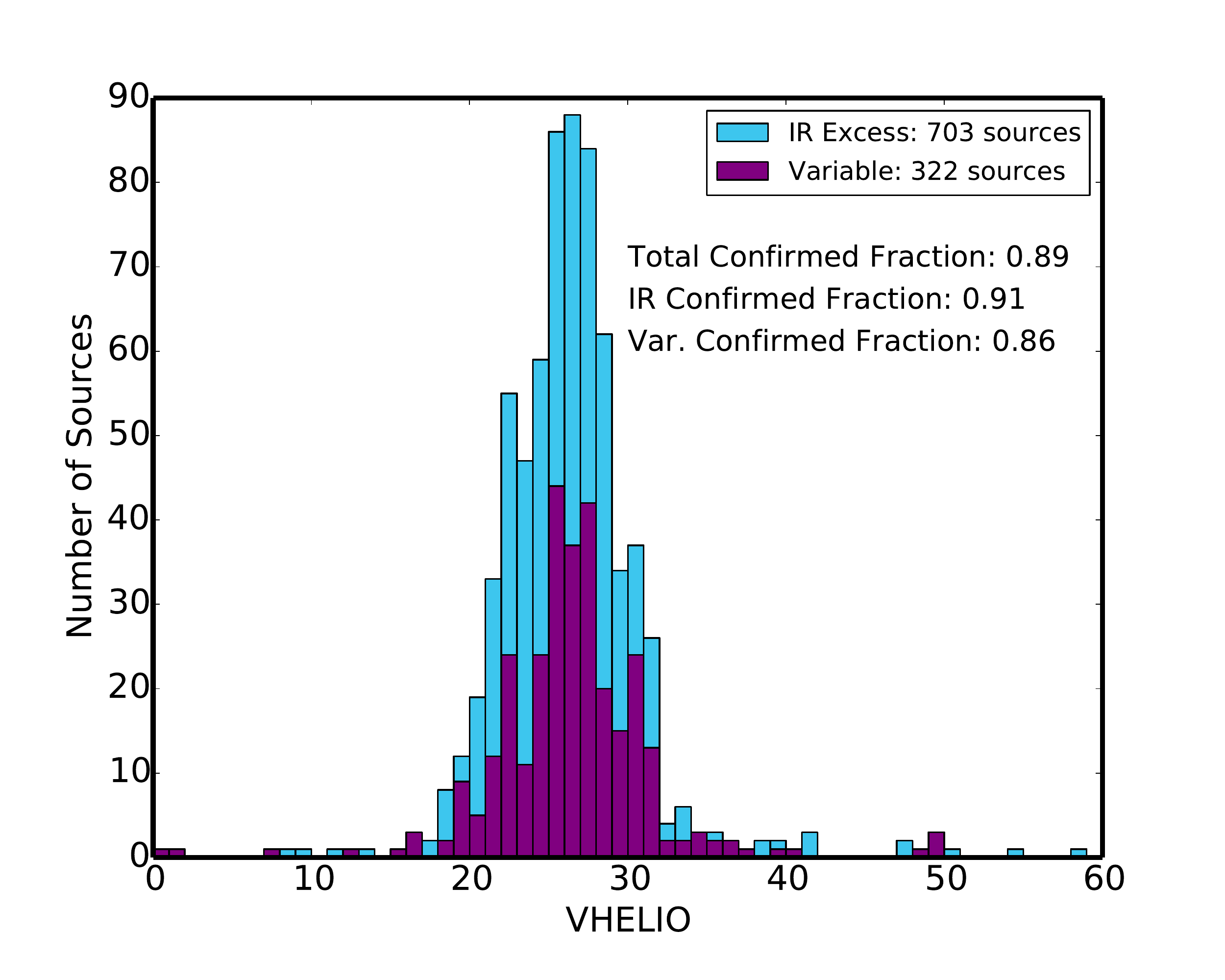}}
\caption{Radial velocities of uniformly selected sources with APOGEE spectra obtained during or before March 2016. Sources include YSOs in all Orion's sub-regions: Orion A \citep[as previously reported by][]{DaRio2016, DaRio2017}, Orion B, OrionOB1a/b and Lambda Ori. Sources included in the uniformly selected sample are shown in two histograms, with 1 km/s bins: the RV distribution of sources selected due to an evidence of circumstellar disk through IR excess are shown in cyan, while sources flagged as optically variable are shown in purple. The RV distribution of variability selected sources is less centrally peaked, indicating a lower return of high-likelihood kinematic members for this selection method. }
\label{fig:vHelioIRVarMatches}
\end{figure*}

To investigate the yield of our uniformly selected candidate YSOs, we examined RVs measured by the APOGEE pipeline for all uniformly selected candidate YSOs with APOGEE-1 or APOGEE-2 spectra observed before April 2016. %For sources in the Orion A sub-region we used the heliocentric velocities determined for the IN-SYNC program. 
Of the 815 uniformly selected candidate YSOs with APOGEE spectra, 210 exhibited an IR excess and optical variability; 493 exhibited only an IR excess, and 112 exhibited only optical variability. The majority (∼70\%) of these spectra were originally observed as part of the IN-SYNC survey of Orion A \citep{DaRio2016,DaRio2017}, but the sources they represent meet the criteria for inclusion in the uniformly selected sample presented in this work. The remaining 30\% of this first set of spectra of uniformly selected targets consists of YSOs in the Lambda Ori, Orion OB1ab, and Orion B regions that were observed in Winter 2015-2016 as part of the APOGEE-2 YSO Goal Science Program.  Only 20\% of the 2700 YSOs targeted by \citet{DaRio2016} in Orion A were selected for inclusion in the uniformly selected sample, however, demonstrating that the uniformly selected sample's strength will be in providing a homogeneous and representative, but not complete, sample of the stellar populations that constitute the OSFC.  Many of the 80\% of the \citet{DaRio2016} targets that are not included in the uniformly selected sample are still included in the APOGEE-2 target list, however, as candidates from the literature were simply included at a lower priority than the uniformly selected YSO candidates.  

Comparing the heliocentric RVs measured by the APOGEE pipeline for the uniformly selected sources to the mean velocities previously measured for key sub-populations in Orion provides support for the fidelity of our target selection methods. As shown in Figure \ref{fig:vHelioIRVarMatches}, the distribution of RVs measured for APOGEE observed sources is centered at $\sim 25$ km/s, with sharp boundaries at $\sim$20 and $\sim$32 km/s. This distribution agrees well with the central velocities previously measured for sub-regions in Orion, which range from 24.5 km/s in $\lambda$ Ori \citep{Dolan2001} to 31 km/s in $\sigma$ Ori \citep{jeffries2006}. Notably, there are are also very few sources at RVs $<$20 and $>$32 km/sec,  where Figure \ref{fig:vrad_besancon} indicates we should expect to see a fair fraction of any Galactic thin and thick disk contaminants in the uniformly selected sample. To more quantitatively evaluate the yield of our uniformly targeted sample of candidate YSOs, we consider sources with heliocentric velocities between 20 km/s and 35 km/s as `RV-confirmed YSO candidates'. We note that final confirmation as a genuine YSO requires a more detailed consideration of the source's spectroscopically determined stellar parameters, which we defer to future papers analyzing this sample. %The distribution of sources shown in Fig. \ref{fig:vHelioIRVarMatches}, drops off significantly past these boundaries. 
Of the current sample of APOGEE-observed uniformly selected sources, 89\% meet this definition of an RV-confirmed YSO. The numbers of RV-confirmed candidates by field are shown in Table 4. %\ref{tab:RVSummary_APOGEE-2_OrionFields}
Only 11\% of the uniformly selected sample exhibit RVs more consistent with membership in the Galactic thin and thick disk than the OSFC; further examination of the sample to remove the lowest confidence RV measurements may further reduce the fraction of genuine contaminants. 
 
\begin{deluxetable}{ccc}
\tablewidth{\columnwidth}
\tabletypesize{\scriptsize}
\tablecaption{RV confirmation for uniformly selected candidates by field}
\label{tab:RVSummary_APOGEE-2_OrionFields}
\tablehead{ 
\colhead{Field} & \colhead{Number} & \colhead{Number} \\
\colhead{Name} & \colhead{Confirmed} & \colhead{Rejected}
}
\startdata
$\lambda$ Ori A & 69 & 14  \\
Orion B & 120 & 26 \\
OB1-ab & 7 & 6 \\
Orion A & 531 & 26\\
\enddata

\end{deluxetable}

%\section{Discussion}

\section{Conclusions} \label{sec:conclusions}

We have designed, implemented, and validated the targeting algorithms for the APOGEE-2 survey of the Orion Star Forming Complex (OSFC). The overarching strategy of the targeting effort is to maximize the yield of bona fide cluster members, while preserving a sub-sample whose simple selection function is optimized for comparative analyses across the entire complex.  The uniformly selected stars provide this latter subset, and were prioritized in targeting to ensure they did not need to `compete' with other target classes with more complex selection biases. The remainder of the targeted sample is subject to a highly heterogeneous set of selection effects, but the expanded sample will be useful for analyses that are less sensitive to those effects -- e.g., measurements of the chemical composition within a spatially compact region, where the selection function is effectively simpler.

In detail, we have:

\begin{enumerate}
\item{applied the criteria developed by KL14 to identify YSOs with warm circumstellar material throughout the OSFC. Using catalogs of previously identified YSOs in Taurus and Orion, we validated the criteria's ability to identify YSOs with 2-24 $\micron$ IR excesses. Applying these cuts to the full catalog of 2MASS+WISE sources across the 420 square degree OSFC, we identify 2699 likely YSOs with IR-excesses.  Of these, 1307 are brighter than the $H\leq$12.4 faint limit we adopt for the APOGEE-2 Orion Survey.}
\item{developed new criteria to identify candidate YSOs on the basis of elevated photometric variability in multi-epoch optical Pan-STARRS photometry.  We demonstrate that about half of all known YSOs in Orion exhibit $>$3$\sigma$ optical variability in a single Pan-STARRS filter, and requiring $\geq$3$\sigma$ variability in multiple Pan-STARRS bands further biases the resultant sample towards bona fide YSOs.  We identify 3697 optical variables throughout the OSFC with $\geq$3$\sigma$ variability in at least three filters, and mean positions in the $g-H$ vs. $g$ and $r-H$ vs. $r$ color-magnitude diagrams consistent with membership in the OSFC. Of these, 990 are brighter than $H\leq$12.4.}
\item{merged the uniformly selected IR-excess and optically variable candidate YSOs with bona fide members drawn from previous studies of Orion's sub-populations to develop plate designs for 16 APOGEE fields spanning the OSFC.  Using a prioritization scheme to assign fibers to target in this merged catalog, we developed 57 distinct plate designs to resolve as many fiber collisions as possible, specially in crowded regions. In total, we targeted 7069 targets, including 1246 uniformly selected YSO candidates and nearly 2500 previously confirmed members.}
\item{analyzed RVs determined by the APOGEE-2 pipeline for 815 uniformly selected targets within 5 APOGEE-2 Orion fields observed in twelve distinct plate designs in early 2016, or with APOGEE-1 spectra from the IN-SYNC Orion A survey \citep{DaRio2016}.  Nearly 90\% of these uniformly selected YSO candidates have RVs consistent with membership in Orion; assuming this yield holds for the 431 uniformly selected YSO candidates that remain to be observed, the APOGEE-2 Orion survey will provide precise stellar parameters for a uniform, homogenously selected sample of more than 1100 members of the OSFC. The simple selection function used to create this subset of the larger survey sample will allow robust inferences to be drawn regarding the relative ages and evolutionary states of members across all major sub-regions of this benchmark laboratory for star and cluster formation studies.}
\end{enumerate}

\vspace{-1cm}
\acknowledgements 
We thank Drew Chojnowski for his timely and efficient efforts to convert our final target lists into complete APOGEE-2 plate designs, without which the Winter 2015-2016 observations would likely not exist. We thank Jonathan Tan for helping define the scope and footprint of the APOGEE-2 Orion Survey during the SDSS-IV planning process.  We thank Marina Kounkel and %Jesus Hernandez for detailed advice regarding target selection strategy in the $\lambda$ Ori fields, and 
Mike Skrutskie for feedback that improved the presentation of this work. \\

Authorship Statement: J.I.C. implemented the KL14 algorithm, defined the variability metrics and cuts for the Pan-STARRS data, verified performance of both IR and Variability selections against existing membership catalogs, merged IR Excess and Variability selected catalogs, evaluated the yield of the uniformly selected sample by comparing RV's measured for APOGEE observed sources to known cluster velocities, generated figures and drafted text for the manuscript. K.R.C. co-leads the APOGEE-2 Young Cluster working group, conceived the high level targeting strategy, oversaw the project's progress, contributed to the analysis of the performance and yield of the uniformly selected sample, prioritized and merged target lists to produce final plate designs, and drafted text for the manuscript.
G. Su\'arez led the catalog compilation, target selection, prioritization and general design of the Orion OB1a/b plates. He generated figures and drafted text for corresponding discussions in Section 4.
%C.R.Z. contributed to the organization of the APOGEE-2 Young Cluster working group, conceived the priority bit select scheme and contributed with text and a figure.
C.R.Z. co-leads the APOGEE-2 Young Cluster working group, contributed to defining the high level targeting strategy, implementing target selection by collating previously identified members and selecting new candidate members using photometry from the literature, and drafted and edited significant portions of the manuscript. 
K.G. and E.F. contributed X-ray selected YSOs from the literature and work that is currently in progress.
E.S., K.C., E.M. and J.V. contributed proprietary Pan-STARRS data for the identification of optically variable sources.
J.J.D. contributed to target selection and design of the Orion OB1a/b plates, and lead the radial velocity analysis in Section 5. 
J.H. contributed to the target selection of the $\lambda$ Ori plates described in Section 4.2.2.
G.Z. and N.D.L. finalized plate designs, oversaw the production of aluminum plug plates and integrated the Orion fields into the APOGEE-2 observing schedule.
J.Y., K.S., G.S.S., J.B., J.K., N.D.R and A.L. are members of the APOGEE-2 Young Cluster working group and provided feedback on the high level targeting strategy of the APOGEE-2 Orion survey.
P.M.F., K.P., D.N., and M.S. contributed to the hardware, software and logistical support required to acquire, reduce and analyze APOGEE 2 spectra.
C.N. served as Chilean Lead Observer of the project SDSSIV/APOGEE 2 South throughout the duration of the project.
R.E.M. contributed general reading of the manuscript and checks for consistency in the instrumental and science case. 
KRC and JIC acknowledge support provided by the NSF through grant AST-1449476. 
CRZ acknowledges support from programs UNAM-DGAPA-PAPITT IN116315 and IN108117.
REM acknowledges support by the BASAL Centro de Astrof\'isica y Techolog\'as Afines (CATA) PFB--6/2007.
PMF acknowledges support from NSF Grant AST-11311835
J.B. is supported by the Ministry for the Economy, Development, and Tourism's Programa Inicativa Cient\'{i}fica Milenio through grant IC120009,  awarded to The Millennium Institute of Astrophysics (MAS).

This research has made use of NASA's Astrophysics Data System Bibliographic Services, the SIMBAD database, operated at CDS, Strasbourg, France, and the VizieR catalogue access tool, CDS, Strasbourg, France \citep{Ochsenbein2000}. 

Funding for the Sloan Digital Sky Survey IV has been provided by the Alfred P. Sloan Foundation, the U.S. Department of Energy Office of Science, and the Participating Institutions. SDSS-IV acknowledges support and resources from the Center for High-Performance Computing at the University of Utah. The SDSS web site is www.sdss.org.

SDSS-IV is managed by the Astrophysical Research Consortium for the Participating Institutions of the SDSS Collaboration including the Brazilian Participation Group, the Carnegie Institution for Science, Carnegie Mellon University, the Chilean Participation Group, the French Participation Group, Harvard-Smithsonian Center for Astrophysics, Instituto de Astrof\'isica de Canarias, The Johns Hopkins University, Kavli Institute for the Physics and Mathematics of the Universe (IPMU) / University of Tokyo, Lawrence Berkeley National Laboratory, Leibniz Institut f\"ur Astrophysik Potsdam (AIP), Max-Planck-Institut f\"ur Astronomie (MPIA Heidelberg), Max-Planck-Institut f\"ur Astrophysik (MPA Garching), Max-Planck-Institut f\"ur Extraterrestrische Physik (MPE), National Astronomical Observatory of China, New Mexico State University, New York University, University of Notre Dame, Observat\'ario Nacional / MCTI, The Ohio State University, Pennsylvania State University, Shanghai Astronomical Observatory, United Kingdom Participation Group, Universidad Nacional Aut\'onoma de M\'exico, University of Arizona, University of Colorado Boulder, University of Oxford, University of Portsmouth, University of Utah, University of Virginia, University of Washington, University of Wisconsin, Vanderbilt University, and Yale University.

The Two Micron All Sky Survey was a joint project of the University of Massachusetts and the Infrared Processing and Analysis Center (California Institute of Technology). The University of Massachusetts was responsible for the overall management of the project, the observing facilities and the data acquisition. The Infrared Processing and Analysis Center was responsible for data processing, data distribution and data archiving.

This publication makes use of data products from the Wide-field Infrared Survey Explorer, which is a joint project of the University of California,  Los Angeles, and the Jet Propulsion Laboratory/California Institute of  Technology, and NEOWISE, which is a project of the Jet Propulsion Laboratory/California Institute of Technology. WISE and NEOWISE are funded  by the National Aeronautics and Space Administration.

This research was made possible through the use of the AAVSO Photometric All-Sky Survey (APASS), funded by the Robert Martin Ayers Sciences Fund. 

This research made use of Astropy, a community-developed core Python package for Astronomy (Astropy Collaboration, 2013).

\appendix

%machine readable version: CottleIR_Excess_Final.txt (stored on overleaf)
%----- Table Sample of Koenig Selected Catalog -----

\floattable
\begin{rotatetable*}
\begin{deluxetable*}{cccccccccccccccccc}
\tablewidth{0pt}
\tabletypesize{\scriptsize}
\tablecaption{Koenig Selected \label{tab:IRmembership}}
\tablehead{\colhead{R.A.} & \colhead{Dec.} & \colhead{W1} & \colhead{$\sigma_{W1}$} & \colhead{W2} & \colhead{$\sigma_{W2}$} & \colhead{W3} & \colhead{$\sigma_{W3}$} & \colhead{W4} & \colhead{$\sigma_{W4}$} & \colhead{J} & \colhead{$\sigma_{J}$} & \colhead{H} & \colhead{$\sigma_{H}$} & \colhead{K$_s$} & \colhead{$\sigma_{K_s}$} & \colhead{} \\ 
\colhead{(deg.)} & \colhead{(deg.)} & \colhead{(mag)} & \colhead{(mag)} & \colhead{(mag)} & \colhead{(mag)} & \colhead{(mag)} & \colhead{(mag)} & \colhead{(mag)} & \colhead{(mag)} & \colhead{(mag)} & \colhead{(mag)} & \colhead{(mag)} & \colhead{(mag)} & \colhead{(mag)} & \colhead{(mag)} & \colhead{Class}}
\startdata
88.9043767 & -12.3980204 & 8.08 & 0.03 & 7.25 & 0.02 & 4.91 & 0.01 & 2.74 & 0.01 & 11.94 & 0.03 & 10.71 & 0.02 & 9.66 & 0.02 & II \\
89.0623887 & -12.3666256 & 10.76 & 0.02 & 10.33 & 0.02 & 8.34 & 0.02 & 6.23 & 0.05 & 13.07 & 0.02 & 12.12 & 0.02 & 11.48 & 0.03 & II \\
89.2096355 & -12.2446224 & 11.00 & 0.02 & 10.50 & 0.02 & 8.31 & 0.02 & 6.34 & 0.07 & 12.71 & 0.02 & 11.92 & 0.02 & 11.65 & 0.02 & II \\
81.0287885 & -12.1691326 & 10.04 & 0.02 & 9.12 & 0.02 & 6.06 & 0.01 & 3.62 & 0.02 & 13.26 & 0.05 & 12.39 & 0.05 & 11.52 & 0.04 & I \\
89.7099552 & -12.1563323 & 9.57 & 0.02 & 9.09 & 0.02 & 7.39 & 0.02 & 5.68 & 0.04 & 11.44 & 0.02 & 10.66 & 0.02 & 10.22 & 0.02 & II \\
84.6836122 & -11.2364286 & 8.80 & 0.02 & 8.30 & 0.02 & 5.63 & 0.01 & 3.80 & 0.02 & 10.01 & 0.02 & 9.64 & 0.03 & 9.34 & 0.02 & II \\
86.8330931 & -10.6367774 & 11.64 & 0.02 & 11.32 & 0.02 & 9.93 & 0.05 & 7.79 & 0.16 & 12.87 & 0.02 & 12.24 & 0.03 & 11.94 & 0.02 & II \\
79.0883956 & -10.5615072 & 10.41 & 0.03 & 9.45 & 0.02 & 6.40 & 0.02 & 4.21 & 0.03 & 13.19 & 0.05 & 12.21 & 0.04 & 11.24 & 0.03 & I \\
88.3544222 & -10.4583884 & 10.13 & 0.02 & 9.35 & 0.02 & 6.92 & 0.02 & 4.80 & 0.03 & 13.13 & 0.02 & 11.78 & 0.02 & 10.97 & 0.02 & II \\
82.7408965 & -10.4430291 & 8.94 & 0.02 & 8.36 & 0.02 & 5.98 & 0.01 & 3.75 & 0.02 & 11.66 & 0.03 & 10.53 & 0.03 & 9.72 & 0.03 & II \\
%\nodata & \nodata & \nodata & \nodata & \nodata & \nodata & \nodata & \nodata & \nodata & \nodata & \nodata & \nodata & \nodata & \nodata & \nodata & \nodata & \nodata & \nodata \\
\enddata
\tablecomments{The full table is published electronically; we show the first 10 lines here as an example of the table's content and format.}
\end{deluxetable*}
\end{rotatetable*}

\clearpage

\begin{rotatetable*}
\begin{deluxetable*}{ccccccccccccccccccc}
\tabletypesize{\tiny}
\tablecaption{Optical Variables in the OSFC \label{tab:PSmembership2}}
\tablehead{\colhead{R.A.} & \colhead{Dec.} & \colhead{} & \colhead{} & \colhead{$\sigma_g$} & \colhead{err$_{g}$} & \colhead{} & \colhead{$\sigma_r$} & \colhead{err$_{r}$} & \colhead{} & \colhead{$\sigma_i$} & \colhead{err$_{i}$} & \colhead{} & \colhead{$\sigma_z$} & \colhead{err$_{z}$} & \colhead{} & \colhead{$\sigma_y$} & \colhead{err$_{y}$} & \colhead{} \\ 
\colhead{(deg.)} & \colhead{(deg.)} & \colhead{N$_{var}$} & \colhead{$\zeta_g$} & \colhead{(Jy)} & \colhead{(Jy)} & \colhead{$\zeta_r$} & \colhead{(Jy)} & \colhead{(Jy)} & \colhead{$\zeta_i$} & \colhead{(Jy)} & \colhead{(Jy)} & \colhead{$\zeta_z$} & \colhead{(Jy)} & \colhead{(Jy)} & \colhead{$\zeta_y$} & \colhead{(Jy)} & \colhead{(Jy)} & \colhead{N$_{obs}$}}
\startdata
86.974592 & -12.44002 & 4 & 0.58 & 1.08e-04 & 5.41e-06 & 0.91 & 4.66e-04 & 1.08e-05 & 0.98 & 2.32e-03 & 4.62e-05 & 1.01 & 5.08e-03 & 9.36e-05 & 0.88 & 4.03e-03 & 1.01e-04 & 28 \\
88.904382 & -12.398028 & 5 & 1.31 & 1.38e-03 & 1.12e-05 & 1.51 & 2.57e-03 & 1.30e-05 & 1.08 & 1.39e-03 & 1.89e-05 & 0.84 & 1.79e-03 & 4.26e-05 & 0.98 & 2.58e-03 & 4.42e-05 & 37 \\
89.06237 & -12.366663 & 4 & 0.28 & 2.78e-05 & 2.12e-06 & 1.09 & 3.34e-04 & 3.91e-06 & 0.89 & 3.37e-04 & 6.25e-06 & 0.81 & 4.75e-04 & 1.07e-05 & 0.98 & 1.09e-03 & 1.63e-05 & 48 \\
79.192648 & -12.349896 & 4 & 1.03 & 1.41e-04 & 2.24e-06 & 0.57 & 1.54e-04 & 7.12e-06 & 1.05 & 4.56e-04 & 6.99e-06 & 0.88 & 3.31e-04 & 7.52e-06 & 1.05 & 6.19e-04 & 9.48e-06 & 34 \\
86.615571 & -12.284184 & 3 & -0.24 & 3.99e-05 & 1.23e-05 & 0.84 & 1.29e-03 & 3.32e-05 & -0.99 & 3.06e-05 & 5.34e-05 & 0.91 & 1.86e-03 & 4.01e-05 & 0.77 & 1.63e-03 & 4.91e-05 & 32 \\
88.03985 & -12.280015 & 5 & 0.73 & 1.64e-04 & 5.04e-06 & 0.86 & 5.83e-04 & 1.33e-05 & 0.79 & 5.01e-04 & 1.35e-05 & 0.90 & 1.19e-03 & 2.45e-05 & 0.93 & 1.53e-03 & 2.97e-05 & 37 \\
80.172356 & -12.252158 & 5 & 1.04 & 1.56e-03 & 3.45e-05 & 1.24 & 3.04e-03 & 4.28e-05 & 1.00 & 3.12e-03 & 7.55e-05 & 0.85 & 2.43e-03 & 8.29e-05 & 1.02 & 3.35e-03 & 7.74e-05 & 17 \\
85.063061 & -12.116306 & 5 & 0.83 & 1.35e-04 & 3.62e-06 & 0.88 & 3.42e-04 & 8.05e-06 & 0.94 & 3.53e-04 & 7.24e-06 & 0.62 & 1.68e-04 & 7.25e-06 & 0.94 & 5.71e-04 & 1.18e-05 & 31 \\
78.411107 & -11.85181 & 3 & -0.08 & 1.51e-05 & 2.47e-06 & -0.12 & 4.87e-05 & 8.68e-06 & 0.67 & 4.78e-04 & 1.38e-05 & 1.21 & 2.44e-03 & 2.02e-05 & 0.76 & 1.23e-03 & 2.85e-05 & 55 \\
76.245795 & -11.796822 & 3 & 0.58 & 1.21e-04 & 5.08e-06 & 0.39 & 1.91e-04 & 1.25e-05 & 0.81 & 5.98e-04 & 1.48e-05 & 1.23 & 1.42e-03 & 1.34e-05 & 0.92 & 1.14e-03 & 2.21e-05 & 39 \\
\enddata
\tablecomments{Targets selected from optical variability. The full table is published electronically, we show the first 10 lines here as an example of the table's content and format. All flux measurements are reported in units of Janskys.}
\end{deluxetable*}
\end{rotatetable*}

%--------------------------BIBLIOGRAPHY---------------------------
\clearpage

\setlength{\baselineskip}{0.6\baselineskip}
\bibliographystyle{aasjournal.bst}

%\bibliography{references}
\setlength{\baselineskip}{1.667\baselineskip}

\end{document}